\documentclass[journal]{IEEEtran}
\usepackage{amsmath,amsfonts,amssymb}
\usepackage{algorithmic}
\usepackage{algorithm}
\usepackage{array}
\usepackage[caption=false,font=normalsize,labelfont=sf,textfont=sf]{subfig}
\usepackage{textcomp}
\usepackage{stfloats}
\usepackage{url}
\usepackage{verbatim}
\usepackage{graphicx}
\usepackage{cite}
\usepackage{adjustbox}
\usepackage{xcolor}

\newtheorem{proposition}{Proposition}

\allowdisplaybreaks

\begin{document}

\title{Markov Chain Monte Carlo Multi-Scan\\ Data Association for Sets of Trajectories}

\author{Yuxuan Xia, {\'A}ngel F. Garc{\'i}a-Fern{\'a}ndez, and Lennart Svensson
\thanks{This work was partially supported by the Wallenberg AI, Autonomous Systems and Software Program (WASP) funded by the Knut and Alice Wallenberg Foundation.
  
Yuxuan Xia is with Zenseact AB, 41756 Gothenburg, Sweden, and the Department of Electrical Engineering, Link{\"o}ping University, 58183 Link{\"o}ping, Sweden (e-mail: yuxuan.xia@zenseact.com, yuxuan.xia@liu.se). 

Lennart Svensson is with the Department of Electrical Engineering, Chalmers University of Technology, 41296 Gothenburg, Sweden (e-mail: lennart.svensson@chalmers.se). 

{\'A}ngel F. Garc{\'i}a-Fern{\'a}ndez is with the Department of Electrical Engineering and Electronics, University of Liverpool, L69 3BX Liverpool, U.K., and also with the ARIES research centre, Universidad Antonio de Nebrija, Madrid, Spain (e-mail: angel.garcia-fernandez@liverpool.ac.uk). 
}
}

\maketitle

\begin{abstract}
This paper considers a batch solution to the multi-object tracking problem based on sets of trajectories. Specifically, we present two offline implementations of the trajectory Poisson multi-Bernoulli mixture (TPMBM) filter for batch data based on Markov chain Monte Carlo (MCMC) sampling of the data association hypotheses. In contrast to online TPMBM implementations, the proposed offline implementations solve a large-scale, multi-scan data association problem across the entire time interval of interest, and therefore they can fully exploit all the measurement information available. Furthermore, by leveraging the efficient hypothesis structure of TPMBM filters, the proposed implementations compare favorably with other MCMC-based multi-object tracking algorithms. Simulation results show that the TPMBM implementation using the Metropolis-Hastings algorithm presents state-of-the-art multiple trajectory estimation performance.
\end{abstract}

\begin{IEEEkeywords}
Multiple object tracking, data association, sets of trajectories, smoothing, Markov chain Monte Carlo.
\end{IEEEkeywords}

\section{Introduction}

Multi-object tracking (MOT) is about the joint estimation of the number of objects and the object trajectories from sensor measurements corrupted with noise \cite{challa2011fundamentals,meyer2018message,streit2021analytic}. Previous works on MOT with trajectory estimation include vector-type methods and set-type methods using labels. The former describes the multi-object states by vectors and constructs trajectories by linking an object estimate with a previous one, while the latter forms trajectories by linking object estimates with the same label. Most of these works focus on the task of estimating the current object states based on all previous measurements, and the main challenge is to handle the data association problem, i.e. to determine the correspondences between measurements and objects.

In this paper, our focus is on a batch formulation of MOT by exploiting all the measurement information in the entire time interval. The fact that offline algorithms generally do not have to be as computationally efficient as online algorithms enables us to revisit and update data associations at previous time steps in light of new observations and thereby improve the estimation of previous object states. These offline algorithms are important for trajectory analytics, with various applications in, e.g. cell tracking, sports athletes tracking, and training data collection for learning-based MOT algorithms. 

The main challenge in batch MOT implementations is to handle the data associations across the entire time interval of interest, where each such global data association can be interpreted as a partition of the sequence of sets of measurements into clutter and clusters of measurements from different time instances originating from the same object. Finding the most likely global data association is a large-scale multidimensional assignment problem. Conventional solutions to this multi-scan data association problem mainly reply on binary programming solvers based on Lagrangian relaxation \cite{popp2001m,poore2006some}. However, these methods can generally only handle data associations up to a few time steps, and thus they typically consider sliding window implementations.

\subsection{Batch MOT with Markov Chain Monte Carlo Sampling}

In this paper, we address this large scale multi-scan data association problem using Markov chain Monte Carlo (MCMC) sampling techniques. In the literature, several MCMC methods have been developed for MOT with image sequences where the proposal distributions for drawing samples of data associations are tailored to image detections \cite{hue2002tracking,khan2006mcmc,cruaciun2015stochastic,jiang2018tracking}. Different from these works, we focus on MOT with the standard multi-object models with Poisson birth model for point objects \cite{mahler2007statistical}. This means that each global data association represents a measurement partition where each cluster of measurements originating from the same object is a collection of at most one measurement at different time steps.

An early work using MCMC sampling to address the multi-scan data association problem for point objects is \cite{oh2009markov}, which applies the Metropolis-Hastings (MH) algorithm to iteratively sample data associations from well-designed proposal distributions, incorporated with certain domain-specific knowledge. The MCMC sampling method in \cite{oh2009markov} has later been combined with a particle Gibbs algorithm in \cite{jiang2015bayesian} for model parameter estimation. The problem of MOT with modelling uncertainties was also addressed in \cite{houssineau2021uncertainty} using MH sampling and the possibility theory \cite{dubois2015possibility}. The MCMC sampling method in \cite{oh2009markov} has also been extended in \cite{vu2014particle} to consider a particle filter-based implementation with more sophisticated proposal distributions. Recently, another MOT batch solution has been presented in \cite{vo2019multi}, where the multi-scan data association problem is tackled using Gibbs sampling \cite{vo2019multi}. 

In the above works, conditioned on a global data association hypothesis, objects have deterministic existences in the multi-object density representation. Therefore, object birth and death events need to be sampled to determine trajectory start and end times, resulting in a significantly increased sampling space. In principle, in batch MOT the objective is to compute the multi-object posterior density distribution of the set of trajectories, which captures all the information about the object trajectories. Conditioned on a specific global data association, we would like to reason about the existence probability of each detected trajectory and the probabilities that the trajectory starts and ends at certain time steps. This can be achieved by considering MOT with sets of trajectories \cite{garcia2019multiple}.

\subsection{Trajectory Poisson Multi-Bernoulli Mixtures for MOT}

A variety of multi-object trackers built upon sets of trajectories has been developed \cite{garcia2019trajectory,granstrom2018poisson,garcia2020trajectory,wei2022trajectory,garcia2019multiple,xia2019multi,xia2019extended,garcia2021continuous,xia2022trajectory,xia2022multiple,garcia2022tracking,zhang2023trajectory,garcia2023trajectory}, showing promising performance on various tasks. The analytic solution to MOT for standard multi-object models with point objects and Poisson point process (PPP) birth \cite{mahler2007statistical} is given by the Poisson multi-Bernoulli mixture (PMBM) filter \cite{williams2015marginal,garcia2018poisson}, and its extension to sets of trajectories is the trajectory PMBM filter \cite{granstrom2018poisson}. The trajectory PMBM (TPMBM) filter has achieved state-of-the-art trajectory estimation performance for both point objects \cite{granstrom2018poisson,garcia2020trajectory} and extended objects \cite{xia2019extended,xia2022trajectory}. 

The trajectory PMBM (TPMBM) filter enjoys an efficient multi-object density representation via the probabilistic existence of each detected trajectory and its start/end times. Importantly, in the TPMBM posterior we have exactly one global hypothesis for every possible partition of the measurements. As a comparison, for multi-object density representations that employ deterministic object existences, the number of global hypotheses also exponentially increases with the number of potential objects to account for their existence uncertainties, resulting in a less efficient hypothesis structure \cite{garcia2018poisson}. 

Furthermore, by modelling the set of undetected trajectories, which are only hypothesized to exist, using a PPP in TPMBM, one is able to reason about the start time and states of a trajectory before it was detected for the first time. The strategy to model undetected objects in PMBM follows naturally from the MOT model assumptions and the PPP birth, but undetected objects are still often ignored by other algorithms that also use a PPP birth, e.g. \cite{oh2009markov,vu2014particle}. Another popular MOT birth model is based on the multi-Bernoulli (MB) process, which is used in \cite{vo2019multi} to implicitly model undetected objects. However, the Bayesian MB birth model is not measurement-driven, and it sets a maximum number of newborn objects at each time, which may not be suitable for many scenarios. Moreover, the TPMBM filtering recursions can be easily adapted to the MB birth model by setting the Poisson intensity for undetected trajectories to zero and adding Bernoulli components of the MB birth instead of the PPP birth in the prediction step \cite{xia2019multi}.

In TPMBM filtering, we can improve past state estimates in the trajectories via smoothing-while-filtering \cite{granstrom2018poisson,garcia2020trajectory}. In principle, the TPMBM filters can compute the exact multi-object posterior of sets of trajectories if they are implemented without approximations. However, in practice, we have to resort to some approximation methods, such as pruning, to keep the computational complexity of the TPMBM implementation at a tractable level. This means that the performance of an online TPMBM filter implementation may decrease when there are too many feasible global data association hypotheses that cannot be effectively enumerated. This motivates us to consider batch TPMBM implementations that can address the large-scale multi-scan data association problem using MCMC sampling techniques.

\subsection{Contributions}

Thanks to the efficient data association hypothesis structure of PMBMs, a PMBM posterior distribution can represent the multi-object posterior distribution with considerably fewer data association hypotheses than multi-object filters that compute a posterior with deterministic object existence \cite{garcia2018poisson}. This is beneficial from an MCMC sampling standpoint as the size of the sampling space is decreased. The existing MOT algorithms that use MCMC sampling for data association \cite{oh2009markov,vu2014particle,vo2019multi} do not have a data association hypothesis structure that is as efficient as a PMBM, and they do not consider the posterior densities on sets of trajectories. This paper aims at bridging this gap in the literature, and its specific contributions include:
\begin{enumerate}
  \item We present two batch TPMBM implementations based on MCMC sampling of data associations using Gibbs sampling and Metropolis-Hastings (MH) sampling, respectively, for standard multi-object models with Poisson birth \cite{mahler2007statistical}. 
  \item In the TPMBM implementation using Gibbs sampling, we adopt a blocked Gibbs sampling strategy to jointly sample a group of data association variables at each time step, making it suitable for Poisson birth. We also show that the conditional distribution in Gibbs sampling can be evaluated in a computationally efficient manner.
  \item In the TPMBM implementation using MH sampling, we show that simple yet flexible proposal distributions are sufficient for sampling the global data association variables. This is possible since uncertainties on trajectory existence and start/end times are captured in the trajectory Bernoulli densities. We also leverage Gibbs sampling when designing the track update move.
  \item The proposed batch TPMBM implementations are compared to several state-of-the-art multi-object smoothing methods. This comparative analysis is conducted in a challenging scenario where objects move in proximity, and it includes an extensive ablation study. The results show that TPMBM implementation using MH sampling yields the best multi-trajectory estimation performance. 
\end{enumerate}

The structure of the rest of the paper is outlined as follows. Models and concepts on trajectories are introduced in Section II. We present the TPMBM filtering recursions in Section III and formulate the batch MOT problem using TPMBM in Section IV. Section V presents the batch TPMBM implementations using MCMC sampling. Simulation results are presented in Section VI, and the paper is concluded in Section VII.

\section{Background}

Let $x_k \in \mathbb{R}^{n_x}$ denote a single object state with dimension $n_x$ at time $k$, and the multi-object state at time $k$ is a set of object states $\mathbf{x}_k \in \mathcal{F}(\mathbb{R}^{n_x})$, where $\mathcal{F}(\mathbb{R}^{n_x})$ denotes the set of finite subsets of space $\mathbb{R}^{n_x}$. Let $z_k \in \mathbb{R}^{n_z}$ denote a single measurement with dimension $n_z$ at time $k$. The set of $m_k$ measurements at time $k$ is $\mathbf{z}_k = \{z_k^1,\dots,z_k^{m_k}\} \in \mathcal{F}(\mathbb{R}^{n_z})$, and the sequence of sets of measurement up to and including time $k$ is $\mathbf{z}_{1:k} = (\mathbf{z}_1,\dots,\mathbf{z}_k)$. 

The cardinality of set $\mathbf{x}$ is denoted as $|\mathbf{x}|$. The Dirac and Kronecker delta functions centred at $x$ are represented using $\delta_x(\cdot)$ and $\delta_x[\cdot]$, respectively. The indicator function for set $D$ is denoted by $1_D(\cdot)$, and the inner product $\int f(x)g(x) dx$ between functions $f(x)$ and $g(x)$ is denoted by $\langle f,g \rangle$. The operation that takes the union of disjoint sets is denoted by $\uplus$.

\subsection{Trajectory Representation}

A trajectory is a finite sequence of single object states at consecutive time steps. A single trajectory can be denoted as $X = (\beta,\varepsilon,x_{\beta:\varepsilon})$, where $\beta$ is the trajectory start time, $\varepsilon$ is the trajectory end time, and $x_{\beta:\varepsilon}$ is the sequence of object states $(x_\beta,x_{\beta+1},\dots,x_{\varepsilon-1},x_\varepsilon)$ with length $\varepsilon-\beta+1$ \cite{granstrom2018poisson}. The space for trajectories up to and including time $k$ is
\begin{equation*}
  \mathcal{T}_{k} = \uplus_{(\beta,\varepsilon)\in I_{k}}\{\beta\}\times\{\varepsilon\}\times\mathbb{R}^{n_x(\varepsilon-\beta+1)},
\end{equation*}
where $I_{k} = \{(\beta,\varepsilon):1\leq\beta\leq\varepsilon\leq k\}$.

The single trajectory function factorizes as
\begin{equation*}
  p(X) = p(x_{\beta:\varepsilon}|\beta,\varepsilon)p(\beta,\varepsilon),
\end{equation*}
where $p(\beta,\varepsilon)$ is defined on $I_{k}$ for $X\in \mathcal{T}_{k}$. The integral of $p(X)$ is given by
\begin{equation*}
  \int_{\mathcal{T}_{k}} p(X) dX = \sum_{(\beta,\varepsilon)\in I_{k}} p(\beta,\varepsilon) \int p(x_{\beta:\varepsilon}|\beta,\varepsilon) dx_{\beta}\cdots dx_\varepsilon.
\end{equation*}
The trajectory Dirac delta function given trajectories $X=(\beta,\varepsilon,x_{\beta:\varepsilon})$ and $X^\prime = (\beta^\prime,\varepsilon^\prime,x^\prime_{\beta^\prime:\varepsilon^\prime})$ is defined as
\begin{equation*}
  \delta_{X}\left[X^\prime\right] = \delta_{\beta}\left[\beta^\prime\right]\delta_{\varepsilon}\left[\varepsilon^\prime\right]\delta_{x_{\beta:\varepsilon}}\left(x^\prime_{\beta^\prime:\varepsilon^\prime}\right).
\end{equation*}

Trajectory functions are typically represented in a mixture form. Consider a trajectory function with $L$ components
\begin{equation}\label{eq_single_trajectory}
  p(X) = \sum_{l=1}^L\nu^lp^l\left(x_{\beta:\varepsilon}|\beta,\varepsilon\right)\delta_{b^l}\left[\beta\right]\delta_{e^l}\left[\varepsilon\right],
\end{equation}
where the $l$-th component has weight $\nu^l$ and state parameters $(b^l,e^l,p^l(\cdot))$ with $b^l\leq e^l$. If $\sum_{l=1}^L\nu^l = 1$, then $p(X)$ is a trajectory density function. If $\sum_{l=1}^L\nu^l \geq 0$, then $p(X)$ is a trajectory intensity function. For trajectory density function of the form \eqref{eq_single_trajectory}, the probability mass function (pmf) of $(\beta,\varepsilon)$ is 
\begin{equation*}
  p(\beta,\varepsilon) = \int p(X)dx_{\beta}\cdots dx_\varepsilon= \sum_{l=1}^L\nu^l\delta_{b^l}[\beta]\delta_{e^l}[\varepsilon].
\end{equation*}

The set of trajectories in time interval $0:k$ is denoted as $\mathbf{X}_{k} \in \mathcal{F}(\mathcal{T}_{k})$. Densities of and integrations over sets of trajectories are defined similarly as densities of and integrations over sets of objects, see \cite{garcia2019multiple} for their explicit expressions.

\subsection{Multi-Trajectory Models}\label{sec_model}

We focus on the estimation of the set $\mathbf{X}_{k}$ of trajectories of all objects that has traversed the surveillance area from time $0$ up to and including the present time $k$. These encompass trajectories of objects that remain within the area at time $k$ as well as trajectories of objects that were in the area earlier but have exited it by time $k$. We consider the standard multi-object models for point objects with Poisson birth \cite[Sec. 9.2.1]{mahler2007statistical}, which is general and can be used to model, e.g. detection-type measurements collected by surveillance and tracking radars. The corresponding multi-trajectory models \cite{granstrom2018poisson} are given as follows.

Each single trajectory $X=(\beta,\varepsilon,x_{\beta:\varepsilon}) \in \mathbf{X}_{k}$ is detected with probability $p^D_k(X) = p^D(x_\varepsilon)\delta_{k}[\varepsilon]$ and generates a single measurement with density $\ell(z|X) = \ell(z|x_\varepsilon)$, or misdetected with probability $1-p^D_k(X)$. Here, $p^D(x)$ is the state-dependent object detection probability. Clutter measurements are modelled using a PPP with intensity function $\lambda^C(\cdot)$.

Given the set $\mathbf{X}_{k}$ of all trajectories, each trajectory $X = (\beta,\varepsilon,x_{\beta:\varepsilon}) \in \mathbf{X}_{k}$ survives to the next time step with probability $p^S(X) = 1$, with a transition density \cite{garcia2020trajectory}
\begin{align}
  &g_{k+1}\left(\beta^\prime,\varepsilon^\prime,x^\prime_{\beta^\prime:\varepsilon^\prime}|X\right) = \delta_{\beta}\left[\beta^\prime\right]\nonumber\\
  &\times\begin{cases}
    \delta_{\varepsilon}\left[\varepsilon^\prime\right]\delta_{x_{\beta:\varepsilon}}\left(x^\prime_{\beta^\prime:\varepsilon^\prime}\right) & \varepsilon^\prime < k \\
    \delta_{\varepsilon}\left[\varepsilon^\prime\right]\delta_{x_{\beta:\varepsilon}}\left(x^\prime_{\beta^\prime:\varepsilon^\prime}\right)\left(1-p^S(x_\varepsilon)\right) & \varepsilon^\prime = k \\
    \delta_{\varepsilon+1}\left[\varepsilon^\prime\right]\delta_{x_{\beta:\varepsilon}}\left(x^\prime_{\beta^\prime:\varepsilon^\prime-1}\right)p^S(x_\varepsilon)g(x_{\varepsilon^\prime}|x_\varepsilon) & \varepsilon^\prime = k+1,
  \end{cases}\label{eq_transition}
\end{align}
where $g(\cdot|x)$ is the object state transition density and $p^S(x)$ is the object survival probability. The interpretation of $p^S(X) = 1$ is that regardless of whether the object survives, its trajectory remains in the set $\mathbf{X}_{k}$ of trajectories. We note that the pmf of trajectory end time is encapsulated in the single trajectory density via \eqref{eq_transition}.

The set ${\bf X}_{k+1}$ of trajectories is the disjoint union of alive trajectories (present in the area), dead trajectories (not present in the area), and new trajectories. The set of newborn trajectories is a PPP with intensity
\begin{equation}
    \lambda_{k+1}^B(X) = \delta_{k+1}[\beta]\delta_{k+1}[\varepsilon]\lambda^B(x_{k+1}),
\end{equation}
where $\lambda^B(x_{k+1})$ is the PPP intensity of newborn objects. Note that new trajectories born at time  $k+1$ have deterministic start and end time $k+1$.

\section{Trajectory PMBM}

In this section, we introduce TPMBM densities and filtering recursions for the standard multi-trajectory models. 

\subsection{Trajectory PMBM Density}

Given measurements $\mathbf{z}_{1:k}$ and the multi-trajectory models described in Section \ref{sec_model}, the density $f_{k^\prime|k}(\cdot)$ of the set of trajectories up to time $k^\prime\in\{k,k+1\}$, is a PMBM \cite{garcia2018poisson,granstrom2018poisson}, with
\begin{align}
  f_{k^\prime|k}({\bf X}_{k^\prime}) &= \sum_{\mathbf{X}_{k^\prime}^u\uplus\mathbf{X}_{k^\prime}^d=\mathbf{X}_{k^\prime}}f^u_{k^\prime|k}\left(\mathbf{X}_{k^\prime}^u\right)f_{k^\prime|k}^d\left(\mathbf{X}_{k^\prime}^d\right), \label{eq_pmbm} \\
  f^u_{k^\prime|k}\left(\mathbf{X}_{k^\prime}^u\right) &= e^{-\left\langle \lambda^u_{k^\prime|k},1 \right\rangle}\prod_{X\in\mathbf{X}^u_{k^\prime}}\lambda_{k^\prime|k}^u(X), \label{eq_ppp} \\
  f_{k^\prime|k}^d\left(\mathbf{X}_{k^\prime}^d\right) &= \sum_{a\in\mathcal{A}_{k^\prime|k}}w^a_{k^\prime|k}\sum_{\uplus_{l=1}^{n_{k^\prime|k}}\mathbf{X}^l_{k^\prime} = \mathbf{X}_{k^\prime}^d}\prod_{i=1}^{n_{k^\prime|k}}f^{i,a^i}_{k^\prime|k}\left(\mathbf{X}_{k^\prime}^i\right), \label{eq_mbm} \\
  f^{i,a^i}_{k^\prime|k}\left(\mathbf{X}_{k^\prime}^i\right) &= \begin{cases}
    1 - r_{k^\prime|k}^{i,a^i} & \mathbf{X}_{k^\prime}^i = \emptyset \\
    r_{k^\prime|k}^{i,a^i}f^{i,a^i}_{k^\prime|k}(X) & \mathbf{X}_{k^\prime}^i = \{X\} \\
    0 & \text{otherwise},
  \end{cases}
\end{align}
where $f_{k|k}({\bf X}_{k})$ (with $k^\prime = k$) represents the filtering density at time $k$ and $f_{k+1|k}({\bf X}_{k+1})$ (with $k^\prime=k+1$) represents the predicted density at time $k+1$. From \eqref{eq_pmbm}, one can observe that a TPMBM is the union of two independent sets: a trajectory PPP \eqref{eq_ppp} with Poisson intensity $\lambda^u_{k^\prime|k}(\cdot)$ representing undetected trajectories, and a trajectory multi-Bernoulli mixture (MBM) \eqref{eq_mbm} representing detected trajectories. In the PMBM filtering recursion, each measurement creates a new Bernoulli component, and the MBM \eqref{eq_mbm} has $n_{k^\prime|k}$ Bernoulli components at time $k^\prime$. Full details on the TPMBM filtering recursion are provided in \cite{granstrom2018poisson}.

A global (data association) hypothesis can be understood as a partition of the batch measurements $\mathbf{z}_{1:k}$ into at most $n_{k^\prime|k}$ non-empty, mutually disjoint subsets originated from the same potential object \cite[Def. 2]{williams2015marginal}. In \eqref{eq_mbm}, $a = (a^1,\dots,a^{n_{k^\prime|k}})\in \mathcal{A}_{k^\prime|k}$ represents a global hypothesis, where $a^i\in\{1,\dots,h^i_{k^\prime|k}\}$ indexes the local hypothesis for the $i$-th Bernoulli component and $h^i_{k^\prime|k}$ is the number of local hypotheses. The $i$-th Bernoulli component with local hypothesis $a^i$ has weight $w_{k^\prime|k}^{i,a^i}$ and density $f_{k^\prime|k}^{i,a^i}(\cdot)$, parameterized by existence probability $r^{i,a^i}_{k^\prime|k}$ and single trajectory density $f_{k^\prime|k}^{i,a^i}(X)$. The global hypothesis weight satisfies $w^a_{k^\prime|k} \propto \prod_{i=1}^{n_{k^\prime|k}}w_{k^\prime|k}^{i,a^i}$, where normalization is required such that $\sum_{a\in\mathcal{A}_{k^\prime|k}}w^a_{k^\prime|k} = 1$.

\begin{figure}[!t]
  \centering
  \includegraphics[width=\linewidth]{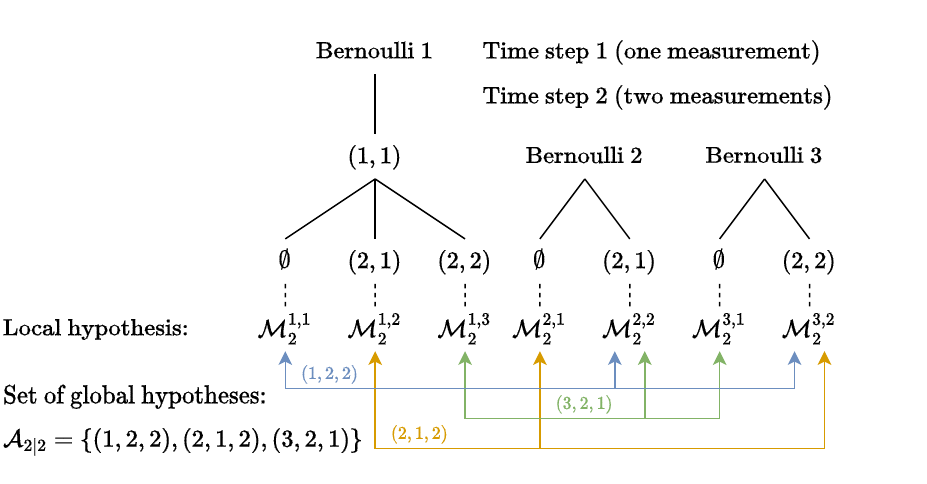}
  \caption{An example of Bernoulli components and hypotheses maintained by a PMBM density of the form \eqref{eq_pmbm}. Since there are three measurements, three Bernoulli components are created in total. At time 2, the existing Bernoulli component (Bernoulli 1)  has three local hypotheses: one for misdetection (marked with $\emptyset$) and two for measurement update (marked with $(t,j)$). Each new Bernoulli component (Bernoulli 2 and 3) has two local hypotheses: one for the case that the measurement corresponds to Bernoulli 1 and one for the case that the measurement is used for updating the PPP. The coloured lines represent three global hypotheses.}
  \label{fig_globalHypo}
  \end{figure}

We proceed to describe the set $\mathcal{A}_{k^\prime|k}$ of global hypotheses. We refer to measurement $z_k^j$, with $j \in \{ 1,\dots,m_k \}$, using $(k,j)$ and denote the set of all such index pairs up to time $k$ by
\begin{equation*}
  \mathcal{M}_k = \{(t,j):t\in \{1,\dots,k\},j\in\{1,\dots,m_t\}\}.
\end{equation*}
Let the set of index pairs correspond to local hypothesis $a^i$ be $\mathcal{M}_k^{i,a^i}\subseteq \mathcal{M}_k$, which can be recursively constructed over time, see, e.g. \cite{williams2015marginal,garcia2018poisson,granstrom2018poisson}. Each global hypothesis must include a local hypothesis for each Bernoulli component and account for the source of each measurement, thus \cite{williams2015marginal}
\begin{multline}\label{eq_set_global_hypothesis}
  \mathcal{A}_{k^\prime|k} = \left\{\left(a^1,\dots,a^{n_{k^\prime|k}}\right): a^i\in \left\{1,\dots,h^i_{k^\prime|k}\right\} \forall i,\right. \\ \left.\biguplus_{i=1}^{n_{k^\prime|k}}\mathcal{M}_k^{i,a^i}  = \mathcal{M}_k,\left| \mathcal{M}_k^{i,a^i} \cap \{(k,j)\}_{j=1}^{m_k} \right| \leq 1, \forall i,a^i \right\}.
\end{multline}
An example of Bernoulli components and hypotheses maintained by PMBM is illustrated in Fig. \ref{fig_globalHypo}.

\subsection{Trajectory PMBM Filtering}

We consider the TPMBM density parameterization where the trajectory Poisson intensity $\lambda^u_{k^\prime|k}(\cdot)$ and the single trajectory density $f_{k^\prime|k}^{i,a^i}(\cdot)$, $i\in\{1,\dots,n_{k^\prime|k}\}$, $a^i\in\{1,\dots,h^i\}$, of each local hypothesis have a mixture representation (cf. \eqref{eq_single_trajectory}),
\begin{align}
  \lambda^u_{k^\prime|k}\left(X\right) &= \sum_{l=1}^{L^u_{k^\prime|k}}\nu_{k^\prime|k}^{u,l}p_{k^\prime|k}^{u,l}\left(x_{\beta:\varepsilon}|\beta,\varepsilon\right)\delta_{b^{u,l}_{k^\prime|k}}[\beta]\delta_{e^{u,l}_{k^\prime|k}}[\varepsilon],\label{eq_traj_poisson}\\
  f_{k^\prime|k}^{i,a^i}\left(X\right) &= \sum_{l=1}^{L^{i,a^i}_{k^\prime|k}}\nu_{k^\prime|k}^{i,a^i,l}p_{k^\prime|k}^{i,a^i,l}\left(x_{\beta:\varepsilon}|\beta,\varepsilon\right)\delta_{b^{i,a^i,l}_{k^\prime|k}}[\beta]\delta_{e^{i,a^i,l}_{k^\prime|k}}[\varepsilon],\label{eq_traj_bernoulli}
\end{align}
where the $l$-th component in single-trajectory function $\lambda^u_{k^\prime|k}(\cdot)$ and $f_{k^\prime|k}^{i,a^i}(\cdot)$ is characterized by $(\nu_{k^\prime|k}^{u,l},p_{k^\prime|k}^{u,l}(\cdot),b^{u,l}_{k^\prime|k},e^{u,l}_{k^\prime|k})$ and $(\nu_{k^\prime|k}^{i,a^i,l},p_{k^\prime|k}^{i,a^i,l}(\cdot),b^{i,a^i,l}_{k^\prime|k},e^{i,a^i,l}_{k^\prime|k})$, respectively.

Since in practice trajectories that were never detected and have ceased to exist are typically considered to be of minimal significance, they do not need to be explicitly represented in practical implementations. That is, the PPP of undetected trajectories here only consider alive trajectories with $e^{u,l}_{k^\prime|k} = k^\prime$, $l\in\{1,\dots, L^u_{k^\prime|k}\}$ in \eqref{eq_traj_poisson}. The Poisson birth intensity $\lambda_{k+1}^B(\cdot)$ is also of the mixture form (cf. \eqref{eq_single_trajectory})
\begin{equation}
  \lambda_{k+1}^B(X) = \delta_{k+1}[\beta]\delta_{k+1}[\varepsilon]\sum_{l=1}^{L^B}\nu^{B,l}p^{B,l}(x_{\beta:\varepsilon}|\beta,\varepsilon).
\end{equation}
Moreover, we use $p_{k^\prime|k}(x_{t})$ to represent the marginalized object density at time $t \in \{\beta,\dots,\varepsilon\}$, obtained from $p_{k^\prime|k}(x_{\beta:\varepsilon})$. 

\begin{proposition}[Trajectory PMBM update]\label{proposition_pmbm_update}
  Assume that the predicted density $f_{k|k-1}(\cdot)$ is a PMBM \eqref{eq_pmbm} with $\lambda^u_{k|k-1}(\cdot)$ and $f^{i,a^i}_{k|k-1}(\cdot)$ given by \eqref{eq_traj_poisson} and \eqref{eq_traj_bernoulli}. The filtering density $f_{k|k}(\cdot)$, updated with measurements $\mathbf{z}_k$, is also a PMBM \eqref{eq_pmbm}.

  For PPP of undetected trajectories, the updated $l$-th component in \eqref{eq_traj_poisson} is
  \begin{subequations}\label{eq_ppp_update}
    \begin{align}
      p_{k|k}^{u,l}(x_{\beta:k}|\beta,k) &= \frac{\left(1-p^D\left(x_{k}\right)\right)p_{k|k-1}^{u,l}(x_{\beta:k}|\beta,k)}{\left\langle1-p^D,p^{u,l}_{k|k-1}\right\rangle},\\
      b^{u,l}_{k|k} & = b^{u,l}_{k|k-1}, \quad e^{u,l}_{k|k} = e^{u,l}_{k|k-1},\\
      \nu_{k|k}^{u,l} &= \left\langle1-p^D,p^{u,l}_{k|k-1}\right\rangle\nu_{k|k-1}^{u,l}.
    \end{align}
  \end{subequations}
  The updated number of Bernoulli components becomes $n_{k|k} = n_{k|k-1} + m_k$. For Bernoulli components that persist from previous time steps $i\in\{1,\dots,n_{k|k-1}\}$, a local hypothesis is included for every possible combination of a local hypothesis from the preceding time step with either a misdetection or an update using one of the $m_k$ new measurements. Consequently, the number of local hypotheses becomes $h^i_{k|k} = h^i_{k|k-1}(1+m_k)$.

  A misdetection hypothesis takes into account both the case that the trajectory does not exist and the case that the trajectory exists but is not present at current time $k$. For misdetection hypotheses with $i\in\{1,\dots,n_{k|k-1}\}$, $a^i\in\{1,\dots,h^i_{k|k-1}\}$:
  \begin{subequations}\label{eq_ber_miss}
    \begin{align}
      &\Phi^{i,a^i,0}_k = \sum_{l=1}^{L^{i,a^i}_{k|k-1}}\delta_k\left[e^{i,a^i,l}_{k|k-1}\right]\nu_{k|k-1}^{i,a^i,l}\left\langle p_{k|k-1}^{i,a^i,l}, p^D \right\rangle, \\
      &w^{i,a^i}_{k|k} = w_{k|k-1}^{i,a^i}\left(1-r_{k|k-1}^{i,a^i}\Phi^{i,a^i,0}_k\right),\\
      &r^{i,a^i}_{k|k} = \frac{r^{i,a^i}_{k|k-1}\left(1-\Phi^{i,a^i,0}_k\right)}{1-r^{i,a^i}_{k|k-1}\Phi^{i,a^i,0}_k},\\
      &p_{k|k}^{i,a^i,l}(x_{\beta:\varepsilon}|\beta,\varepsilon) = \frac{\left(1-p^D(x_k)\right)p_{k|k-1}^{i,a^i,l}(x_{\beta:\varepsilon}|\beta,\varepsilon)}{\left\langle  1-p^D,p_{k|k-1}^{i,a^i,l}\right\rangle},\\
      &b^{i,a^i,l}_{k|k} = b^{i,a^i,l}_{k|k-1}, \quad e^{i,a^i,l}_{k|k} = e^{i,a^i,l}_{k|k-1},\\
      &\nu^{i,a^i,l}_{k|k} = \begin{cases}
        \frac{\nu^{i,a^i,l}_{k|k-1}}{1-\Phi^{i,a^i,0}_k} & e^{i,a^i,l} < k \\
        \frac{\nu^{i,a^i,l}_{k|k-1}\left\langle 1-p^D,p^{i,a^i,l}_{k|k-1} \right\rangle }{1-\Phi^{i,a^i,0}_k} & e^{i,a^i,l} = k.
      \end{cases}
    \end{align}
  \end{subequations}
  
  For hypotheses updating existing Bernoulli components, $i\in\{1,\dots,n_{k|k-1}\}$, $a^i = \widetilde{a}^i + h^i_{k|k-1}j$, $j\in\{1,\dots,m_k\}$, i.e. the previous hypothesis $\widetilde{a}^i$ is updated with measurement $z_k^j$:
  \begin{subequations}\label{eq_ber_update}
    \begin{align}
      &\mathcal{M}_k^{i,a^i} = \mathcal{M}_{k-1}^{i,\widetilde{a}^i} \cup \{(k,j)\},\\
      &\Phi_k^{i,a^i,j} = \sum_{l=1}^{L^{i,\widetilde{a}^i}_{k|k-1}}\delta_k\left[e^{i,\widetilde{a}^i,l}_{k|k-1}\right]\nu_{k|k-1}^{i,\widetilde{a}^i,l}\left\langle p_{k|k-1}^{i,\widetilde{a}^i,l}, \ell\left(z_k^j|\cdot\right)p^D \right\rangle, \\
      &w^{i,a^i}_{k|k} = w^{i,\widetilde{a}^i}_{k|k-1}r^{i,\widetilde{a}^i}_{k|k-1}\Phi_k^{i,a^i,j},\\
      &r^{i,a^i}_{k|k} = 1,\\
      &p^{i,a^i,l}_{k|k}\left(x_{\beta:\varepsilon}|\beta,\varepsilon\right) =\frac {p^D(x_k)p_{k|k-1}^{i,\widetilde{a}^i,l}\left(x_{\beta:\varepsilon}|\beta,\varepsilon\right)\ell\left(z_k^j|x_k\right)}{\left\langle p_{k|k-1}^{i,\widetilde{a}^i,l}, \ell\left(z_k^j|\cdot\right)p^D \right\rangle},\\
      &b^{i,a^i,l}_{k|k} = b^{i,a^i,l}_{k|k-1}, \quad e^{i,a^i,l}_{k|k} = e^{i,a^i,l}_{k|k-1},\\
      &\nu_{k|k}^{i,a^i,l} = \begin{cases}
        0 & e^{i,\widetilde{a}^i,l}_{k|k-1} < k \\
        \frac{\nu_{k|k-1}^{i,\widetilde{a}^i,l}\left\langle p_{k|k-1}^{i,\widetilde{a}^i,l}, \ell\left(z_k^j|\cdot\right)p^D \right\rangle}{\Phi_k^{i,a^i,j}} & e^{i,\widetilde{a}^i,l}_{k|k-1} = k.
      \end{cases}
  \end{align}
  \end{subequations}
  We note that for local hypotheses resulted from measurement update, the trajectory exists and is present at current time $k$ with probability one.

  Finally, each new Bernoulli component, $i = n_{k|k-1} + j$, $j\in \{1,\dots,m_k\}$, has two local hypotheses ($h^i_{k|k} = 2$): 1) measurement $z_k^j$ corresponds to another Bernoulli component; and 2) measurement $z_k^j$ corresponds to either clutter or the first detection of an undetected trajectory:
  \begin{subequations}\label{eq_new}
    \begin{align}
      &\mathcal{M}_k^{i,1} = \emptyset, \quad w_{k|k}^{i,1} = 1, \quad r_{k|k}^{i,1} = 0,\label{eq_nonexist}\\
      &\mathcal{M}_k^{i,2} = \{(k,j)\}, \quad L_{k|k}^{i,2} = L_{k|k-1}^u\\
      &\Phi_k^{u,j} = \sum_{l=1}^{L^u_{k|k-1}}\nu_{k|k-1}^{u,l}\left\langle p_{k|k-1}^{u,l}, \ell\left(z_k^j|\cdot\right)p^D \right\rangle, \\
      &w_{k|k}^{i,2} = \lambda^C\left(z_k^j\right) + \Phi_k^{u,j},\\
      &r_{k|k}^{i,2} = \frac{\Phi_k^{u,j}}{w_{k|k}^{i,2}},\\
      &p_{k|k}^{i,2,l}\left(x_{\beta:k}|\beta,k\right) = \frac{p^D(x_k)p_{k|k-1}^{u,l}\left(x_{\beta:k}|\beta,k\right)\ell\left(z_k^j|x_k\right)}{\left\langle p_{k|k-1}^{u,l}, \ell\left(z_k^j|\cdot\right)p^D \right\rangle},\\
      &b^{i,2,l}_{k|k} = b^{u,l}_{k|k-1}, \quad e^{i,2,l}_{k|k} = e^{u,l}_{k|k-1},\\
      &\nu_{k|k}^{i,2,l} = \frac{\nu_{k|k-1}^{u,l}\left\langle p_{k|k-1}^{u,l}, \ell\left(z_k^j|\cdot\right)p^D \right\rangle}{\Phi_k^{u,j}}.
    \end{align}
  \end{subequations}
\end{proposition}
For the first case, the new Bernoulli component has zero existence probability, whereas for the second case, the probability of existence $r^{i,2}_{k|k}$ models the relative likelihood of $z_k^j$ being clutter or the first detection of an undetected object. We also note that the updated single-trajectory density also captures trajectory information before current time $k$.

\begin{proposition}[Trajectory PMBM prediction]\label{proposition_pmbm_predict} Assume that the filtering density $f_{k|k}(\cdot)$ is a PMBM \eqref{eq_pmbm} with $\lambda^u_{k|k}(\cdot)$ and $f^{i,a^i}_{k|k}(\cdot)$ given by \eqref{eq_traj_poisson} and \eqref{eq_traj_bernoulli}. The predicted density $f_{k+1|k}(\cdot)$ is also a PMBM \eqref{eq_pmbm}. For the PPP of undetected trajectories, we have $L^u_{k+1|k} = L^u_{k|k} + L^B$, and 
  \begin{subequations}\label{eq_ppp_predict}
    \begin{align}
      &p^{u,l}_{k+1|k}\left(x_{\beta:k+1}|\beta,k+1\right) = \frac{p^S(x_k)p^{u,l}_{k|k}\left(x_{\beta:k}|\beta,k\right)g(x_{k+1}|x_k)}{\left\langle p_{k|k}^{u,l},p^S \right\rangle},\\
      &b^{u,l}_{k+1|k} = b^{u,l}_{k|k}, \quad e^{u,l}_{k+1|k} = e^{u,l}_{k|k} + 1,\\
      &\nu^{u,l}_{k+1|k} = \nu_{k|k}^{u,l}\left\langle p_{k|k}^{u,l},p^S \right\rangle, 
    \end{align}
    for alive trajectories with $l\in\{1,\dots,L^u_{k|k}\}$,
    \begin{align}
      &p^{u,l}_{k+1|k}\left(x_{k+1:k+1}|k+1,k+1\right) = p^{B,l-L^u_{k|k}}(x_{k+1}),\\
      &b_{k+1|k}^{u,l} = e_{k+1|k}^{u,l} = k+1,\\
      &\nu_{k+1|k}^{u,l} = \nu^{B,l-L^u_{k|k}},
    \end{align}
    for newborn trajectories with $l\in\{L^u_{k|k}+1,\dots,L^u_{k+1|k}\}$.
  \end{subequations}

  For Bernoulli components describing detected trajectories, we only need to update mixture components with $e^{i,a^i,l}_{k|k} = k$ that represent alive trajectories. Specifically, each such mixture component creates two predicted mixture components, the first of which covers the case that the trajectory ends at time $k$,
  \begin{subequations}\label{eq_ber_predict}
    \begin{align}
      p_{k+1|k}^{i,a^i,l}\left(x_{\beta:k}|\beta,k\right) &= \frac{\left(1-p^S(x_k)\right)p_{k|k}^{i,a^i,l}\left(x_{\beta:k}|\beta,k\right)}{1- \left\langle p_{k|k}^{i,a^i,l}, p^S\right\rangle},\label{eq_end_k}\\
      b^{i,a^i,l}_{k+1|k} &= b^{i,a^i,l}_{k|k}, \quad e^{i,a^i,l}_{k+1|k} = e^{i,a^i,l}_{k|k},\\
      \nu_{k+1|k}^{i,a^i,l} &= \nu_{k|k}^{i,a^i,l}\left(1-\left\langle p_{k|k}^{i,a^i,l}, p^S\right\rangle\right),
    \end{align}
    and the second of which covers the case that the trajectory continues to exist at time $k+1$,
    \begin{align}
      &p_{k+1|k}^{i,a^i,l}\left(x_{\beta:k+1}|\beta,k+1\right) = \frac{p^S(x_k)p_{k|k}^{i,a^i,l}\left(x_{\beta:k}|\beta,k\right)g(x_{k+1}|x_k)}{\left\langle p_{k|k}^{i,a^i,l}, p^S\right\rangle},\label{eq_contiune_k}\\
      &b^{i,a^i,l}_{k+1|k} = b^{i,a^i,l}_{k|k}, \quad e^{i,a^i,l}_{k+1|k} = e^{i,a^i,l}_{k|k} + 1,\\
      &\nu_{k+1|k}^{i,a^i,l} = \nu_{k|k}^{i,a^i,l}\left\langle p_{k|k}^{i,a^i,l}, p^S\right\rangle.
    \end{align}
  \end{subequations}
\end{proposition}
We note that the information on whether the trajectory ends, or if its length is increased by one is compactly encapsulated in the predicted single trajectory density without expanding the global hypothesis space as in \cite{oh2009markov,vu2014particle,vo2019multi}.

\textit{Remark:  Proposition \ref{proposition_pmbm_update} and \ref{proposition_pmbm_predict} include explicit expressions for predicting and updating single trajectory densities of the form specified by (9) and (10), making them more suitable for practical implementations than the TPMBM filtering recursions in \cite{granstrom2018poisson} for general single trajectory density functions.}

\section{Problem Formulation}

We aim at estimating the set $\mathbf{X}_{K}$ of all trajectories through processing the sequence of measurement sets $\mathbf{z}_{1:K}$ in a batch. Formally speaking, our objective is to directly compute the multi-trajectory posterior density $f_{K|K}(\mathbf{X}_{K})$ using  $\mathbf{z}_{1:K}$.

\textit{Remark: Given the multi-object dynamic and measurement models defined for sets of trajectories in Section \ref{sec_model}, if the prior at time $0$ is a TPMBM, then the multi-trajectory posterior $f_{K|K}(\mathbf{X}_{K})$ conditioned on $\mathbf{z}_{1:K}$ is a TPMBM \eqref{eq_pmbm}.}

The main challenge in computing the TPMBM posterior is due to the intractably large number of global hypotheses $|\mathcal{A}_{K|K}|$ in the MBM \eqref{eq_mbm}. In this paper, we tackle this large scale multidimensional data association problem using MCMC sampling. After obtaining samples of global hypotheses from the stationary distribution, we can only keep those with highest weights, which can be understood as minimizing the $L_1$ norm between the MBM and its truncation \cite{granstrom2019poisson}. 

Besides finding the global hypotheses with highest weights, we need to compute the Poisson intensity of undetected trajectories $\lambda^u_{K|K}(\cdot)$ and the local hypothesis densities $f^{i,a^i}_{K|K}(\cdot)$. Specifically, assume that the multi-object prior is $f_{0|0}(\emptyset) = 1$, we can obtain $\lambda^u_{K|K}(\cdot)$ by recursively applying \eqref{eq_ppp_predict} and \eqref{eq_ppp_update}. The number of Bernoulli components is given by the total number of measurements $n_{K|K} = \sum_{k=1}^K m_k$. For Bernoulli component $i\in\{\sum_{t=1}^{k-1}m_t+1,\dots,\sum_{t=1}^{k}m_t\}$ initiated at time $k$, it has $h^i_{K|K} = 1 + \prod_{t = k+1}^K (m_t + 1)$ local hypotheses at time $K$. Local hypothesis density $f^{i,a^i}_{K|K}(\cdot)$, $i\in \{1,\dots,n_{K|K}\}$, $a^i\in\{1,\dots,h^i_{K|K}\}$, with measurement association history $\mathcal{M}_{K|K}^{i,a^i}$ can be computed by recursively applying \eqref{eq_ber_miss}, \eqref{eq_ber_update}, \eqref{eq_new} and \eqref{eq_ber_predict}. We note that for local hypothesis density $f^{i,a^i}_{K|K}(\cdot)$ with $|\mathcal{M}_K^{i,a^i}|\geq 2$, $r_K^{i,a^i} = 1$, and that for $f^{i,a^i}_{K|K}(\cdot)$ with $|\mathcal{M}_K^{i,a^i}| = 1$, $r_K^{i,a^i} \in (0,1)$. This means that we are certain that a Bernoulli component represents an actual trajectory, if and only if it is associated with more than one measurement. 

\section{MCMC Sampling}

In this section, we present techniques for efficiently finding global hypotheses with highest weights in the PMBM posterior using MCMC sampling. Two sampling methods are described, one is based on Gibbs sampling, and the other is based on MH sampling. The relation between the proposed MH sampler and the one in \cite{oh2009markov} is also explained. Finally, we discuss practical considerations in the implementations.

Conceptually, a global data association hypothesis describes a partition of the set $\mathcal{M}_K$ of measurement pairs into clusters of measurement pairs associated to the same local hypothesis. Our objective is to run an MCMC chain to sample such data association hypotheses. However, some representations of how we partition the data into clusters can be mathematically inconvenient to work with. Importantly, for point object tracking, we find it convenient to represent data associations from a track-oriented perspective, in contrast to the measurement-oriented data association used in extended object tracking \cite{xia2022trajectory}, where an object may give rise to more than one measurement at each time step.

To this end, we introduce an alternative track-oriented data association representation to the local hypothesis measurement association history. Specifically, the correspondence between Bernoulli components and measurements at time $k\in\{1,\dots, K\}$ is represented by a vector $\theta_k$ of dimension $n_{k|k}$, where $\theta_k(i)$ denotes the $i$-th entry of vector $\theta_k$. If an existing Bernoulli component $i\in\{1,\dots,n_{k|k-1}\}$ is misdetected or a new Bernoulli component $i\in\{n_{k|k-1}+1,\dots,n_{k|k}\}$ does not exist, then $\theta_k(i) = 0$. If Bernoulli component $i$ corresponds to the $j$-th measurement, then $\theta_k(i) = j$.  

According to how the global hypothesis \eqref{eq_set_global_hypothesis} is defined, there are certain constraints on the construction of the track-oriented data association $\theta_k$, $k\in\{1,\dots, K\}$:
\begin{enumerate}
  \item $\theta_k(i) \in \{0,1,\dots,m_k\},$ $\forall i\in\{1,\dots,n_{k|k-1}\}$.
  \item $\theta_k(i) \in \{0,j\},$ $\forall i = n_{k|k-1}+j$, and $j\in \{1,\dots,m_k\}$.
  \item $\theta_k(i) \neq \theta_k(j)$, $\forall i\neq j, \theta_k(i) > 0$, and $\theta_k(j) > 0$.
  \item $\forall j\in\{1,\dots,m_k\}$, $\exists i\in\{1,\dots,n_{k|k}\}$ such that $\theta_k(i)=j$.
  \item If $\theta_k(i) = 0$, where $i\in \{n_{k|k-1}+1,\dots,n_{k|k}\}$, then $\theta_{k+t}(i) = 0$, $\forall t\in\{1,\dots,K-k\}$.
\end{enumerate}

Constraint 3) and 4) together mean that each measurement should be assigned to exactly one Bernoulli component. Constraint 5) means that if new Bernoulli component $i$, initiated at time $k$, does not exist, then it cannot be associated with any measurements at any subsequent time steps. As a direct consequence of constraint 2) and 5), it also holds that 
\begin{enumerate}
  \item[6)] If $\exists \theta_{k+t}(i) > 0$, where $i\in \{n_{k|k-1}+1,\dots,n_{k|k}\}$ and $t\in\{1,\dots,K-k\}$, then $\theta_k(i) = i-n_{k|k-1}$. 
\end{enumerate}
Constraint 6) means that if Bernoulli component $i$, initiated at time $k$, has been associated with any measurements at subsequent time steps, then it must be created by measurement with index $i-n_{k|k-1}$. 

\begin{figure*}[!t]
  \centering
  \includegraphics[width=0.9\linewidth]{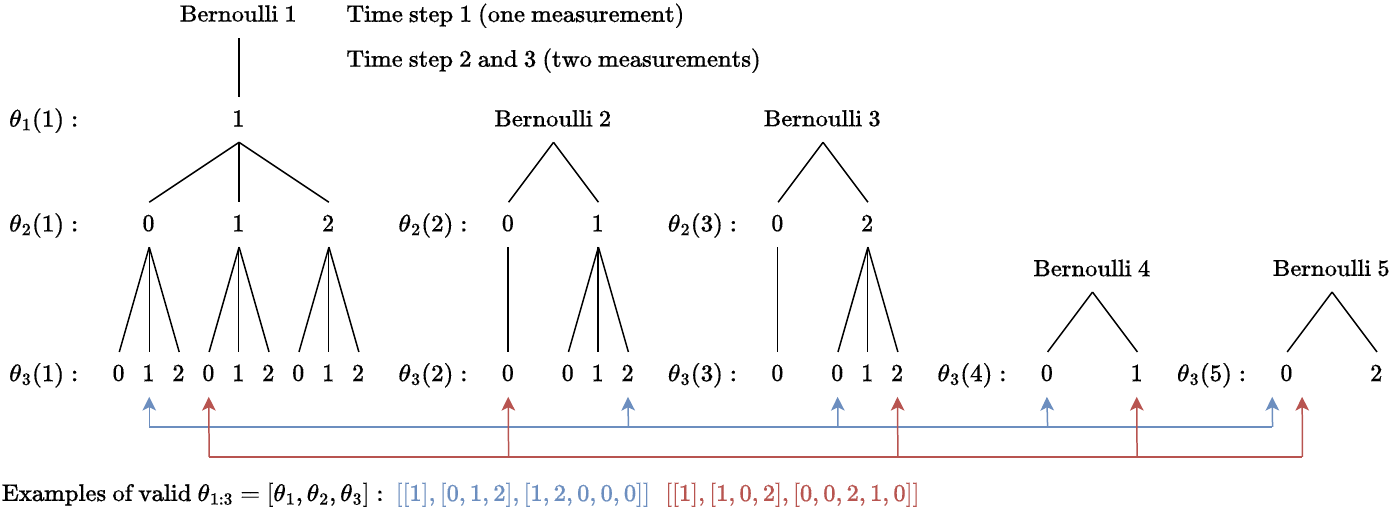}
  \caption{An example of the track-oriented data association sequence $\theta_{1:3}$, where there are 1, 2, 2 measurements at time 1, 2 and 3, respectively. The coloured lines represent two valid $\theta_{1:3}$. For the blue one, Bernoulli $1$ is not detected at time 2 and updated by $z_3^1$ at time 3; Bernoulli 2 is created by $z_2^1$ at time 2 and updated by $z_3^2$ at time 3; Bernoulli $3$ is created by $z_2^2$ at time 2 and not detected at time 3; Bernoulli $4$ and $5$ do not exist. For the red one, Bernoulli 1 is updated by $z_2^1$ at time 2 and not detected at time 3; Bernoulli 3 is created by $z_2^2$ at time 2 and updated by $z_3^2$ at time 3; Bernoulli 4 is created by $z_3^1$ at time 3; Bernoulli 2 and 5 do not exist. The two global hypotheses $a$ correspond to the blue and red $\theta_{1:3}$ are $(2,4,2,1,1)$ and $(4,1,4,2,1)$, respectively.}
  \label{fig_dataAssoc}
\end{figure*}

The collection of all vectors $\theta_k$ at time $k$ is represented by $\Theta_k$, and the time sequence of $\theta_k$ in time interval $1:K$ is $\theta_{1:K} \in \Theta_{1:K}$. With the above constraints 1)--6) in place, each $\theta_{1:k} \in \Theta_{1:k}$ determines a unique partition of $\mathcal{M}_k$. We also note that each legal set of local hypothesis measurement association histories $\{\mathcal{M}_K^{i,a^i}\}$ is a partition of $\mathcal{M}_K$. This also means that there is a bijection between $\Theta_{1:k}$ and set of global hypotheses $\mathcal{A}_{k|k}$, denoted by $\Lambda_k: \Theta_{1:k} \rightarrow \mathcal{A}_{k|k}$. That is, $\Lambda_k(\theta_{1:k}) = a$ means that $\theta_{1:k} \in \Theta_{1:k}$ corresponds to $a\in \mathcal{A}_{k|k}$, and we can use $\Lambda_k^i(\theta_{1:k})$ to represent a local hypothesis $a^i$ in a global hypothesis $a$. An example of the track-oriented data association representation $\theta_{1:K}$ is illustrated in Fig. \ref{fig_dataAssoc}.

Sampling a global hypothesis $a$ from the discrete probability distribution $w^a_{K|K}$, $a\in \mathcal{A}_{K|K}$, is equivalent to sampling 
\begin{align}
  \theta_{1:K} \sim \pi(\theta_{1:K}) &= 1_{\Theta_{1:K}}(\theta_{1:K})w_{K|K}^a \nonumber \\
  &\propto 1_{\Theta_{1:K}}(\theta_{1:K})\prod_{i=1}^{n_{K|K}}w_{K|K}^{i,\theta_{1:K}^i}, \label{eq_full_sample}
\end{align}
where $\pi(\cdot)$ denotes the pmf of $\theta_{1:K}$, and we define $\theta_{1:K}^i \triangleq \Lambda_K^i(\theta_{1:K})$ for notational brevity. In addition, we also write $\theta_k(i:j) \triangleq (\theta_k(i),\dots,\theta_k(j))$.

\subsection{Blocked Gibbs Sampling}\label{sec_gibbs}

To sample from \eqref{eq_full_sample}, we can sequentially draw samples of $\theta_k$, $k\in \{1,\dots,K\}$, from the conditional distribution using Gibbs sampling
\begin{align}
  \pi_{k}\left(\theta_k | \theta_{1:k-1},\theta_{k+1:K}\right) &\propto \pi(\theta_{1:K}) \nonumber \\
  &\propto 1_{\Theta_{1:K}}(\theta_{1:K})\prod_{i=1}^{n_{k|k}}w_{K|K}^{i,\theta_{1:K}^i} \label{eq_condition_sample1}.
\end{align}
We note that the product over $w_{K|K}^{i,\theta_{1:K}^i}$ in \eqref{eq_condition_sample1} goes from $i=1$ to $n_{k|k}$, not $n_{K|K}$ as in \eqref{eq_full_sample}. This proportionality in \eqref{eq_condition_sample1} holds since, conditioned on $\theta_{k+1:K}$, data association $\theta_k(i)$ of Bernoulli component $i\in\{1,\dots,n_{k|k}\}$ at time $k$ does not depend on Bernoulli components created after time $k$. To sample $\theta_k$ from \eqref{eq_condition_sample1}, one possible strategy is to sample one variable $\theta_k(i)$ at a time for $i\in\{1,\dots,n_{k|k}\}$ conditioned on $\theta_{1:k-1}$, $\theta_{k+1:K}$, and $\theta_k(1:i-1)$, $\theta_k(i+1:n_{k|k})$. However, due to the constraints 1)--6) of $\theta_k$, it turns out that $\theta_k(i)$ is a deterministic function of $\theta_k(1:i-1)$ and $\theta_k(i+1:n_{k|k})$. This implies that such a sampling algorithm would never change $\theta_k$.

According to constraint 2), if  measurement $z_k^j$ at time $k$ does not correspond to an existing Bernoulli component $i\in\{1,\dots,n_{k|k-1}\}$, then it must be used to create new Bernoulli component $n_{k|k-1}+j$ where $j\in\{1,\dots,m_k\}$. This means that if we change the data association $\theta_k(i)$ of an existing Bernoulli component $i\in\{1,\dots,n_{k|k-1}\}$ at time $k$ from $j$ to $0$, while fixing the data associations of other existing Bernoulli components, the data association $\theta_k(n_{k|k-1}+j)$ of the $j$-th new Bernoulli component must change from $0$ to $j$, and vice versa. In light of this observation, we adopt a blocked Gibbs sampling strategy to jointly sample the data association $\theta_k(i)$ of an existing Bernoulli component $i\in\{1,\dots,n_{k|k-1}\}$ and the data associations $\theta_k(n_{k|k-1}+1:n_{k|k})$ of all the new Bernoulli components at time $k$ via 
\begin{align}
  \pi_{k,i}^{\theta_k(i)}&\triangleq\pi_k\left(\theta_k(i), \theta_k(n_{k|k-1}+1:n_{k|k}) \right.\nonumber\\ &~~~~~~~~\left. \mid  \theta_{1:k-1},\theta_k(1:i-1), \theta_k(i+1:n_{k|k-1}), \theta_{k+1:K}\right)\nonumber \\ &\propto 1_{\Theta_{1:K}}(\theta_{1:K})w_{K|K}^{i,\theta_{1:K}^i}\prod_{j=1}^{m_k}w_{K|K}^{n_{k|k-1}+j,\theta_{1:K}^{n_{k|k-1}+j}}\nonumber\\
  &\triangleq 1_{\Theta_{1:K}}(\theta_{1:K})w_{K|K}^{i,\theta_{1:K}^i} \Pi_{k,n_{k|k}}^{\theta_k(i)},
  \label{eq_condition_block}
\end{align}
where we introduce the shorthand notations $\pi_{k,i}^{\theta_k(i)}$ and $\Pi_{k,n_{k|k}}^{\theta_k(i)}$. 

\textit{Example: Let us take the red $\theta_{1:3}$ illustrated in Fig. \ref{fig_dataAssoc} for an example, and say we would like to change the data association $\theta_3(3)=2$ of Bernoulli 3 at time 3 while fixing $\theta_3(1)$ and $\theta_3(2)$. If we change $\theta_3(3)$ from 2 to 0, then we must also change $\theta_3(5)$ from 0 to 2.  If we change $\theta_3(3)$ from 2 to 1, then we must also change $\theta_3(4)$ from 1 to 0 and $\theta_3(5)$ from 0 to 2. This means that $\theta_3(3)$ cannot be changed alone while fixing all the other data associations $\theta_3(1:2)$ and $\theta_3(4:5)$. However, it is possible to jointly modify the data associations $\theta_3(3:5)$. That is, we can change $\theta_3(3:5)=[2,1,0]$ to either $[0,1,2]$ or $[1,0,2]$. We also note that the change of $\theta_3(4:5)$ is uniquely determined by $\theta_3(3)$.}

Further, since the data associations $\theta_k(n_{k|k-1}+1:n_{k|k})$ of new Bernoulli components are deterministic functions of the data associations $\theta_k(1:n_{k|k-1})$ of existing Bernoulli components, it is sufficient to only sample the data association $\theta_k(i)$ of existing Bernoulli components for each group of variables $(\theta_k(i),\theta_k(n_{k|k-1}+1:n_{k|k}))$, $i\in\{1,\dots,n_{k|k-1}\}$. Note that this is also why we only explicitly denoted the dependences of $\pi_{k,i}^{\theta_k(i)}$ and $\Pi_{k,n_{k|k}}^{\theta_k(i)}$ on $\theta_k(i)$ for brevity. Moreover, there is a convenient method to sample $\theta_k(i)$ that greatly simplifies the computation of the product of $m_k+1$ factors in \eqref{eq_condition_block}.

The key to simplifying the unnormalized expression in \eqref{eq_condition_block} is to note that the ratio between the conditional probabilities of the two cases $\theta_k(i)=j$, $j\in\{1,\dots,m_k\}$, and $\theta_k(i)=0$ has a simple form. Specifically, the product $\Pi_{k,n_{k|k}}^{\theta_k(i)}$ over the $m_k$ local hypothesis weights in \eqref{eq_condition_block} of the two cases only differs in the $j$-th factor, i.e. the local hypothesis weight that measurement $z_k^j$ initiates a new Bernoulli component, which can take two different values depending on data association $\theta_k(n_{k|k-1}+j)$. First, if $\theta_k(n_{k|k-1}+j)=0$, which means that measurement $z_k^j$ corresponds to an existing Bernoulli component, the corresponding local hypothesis weight is one (cf. \eqref{eq_nonexist}). Second, if $\theta_k(n_{k|k-1}+j)=j$, then measurement $z_k^j$ corresponds to the $j$-th new Bernoulli component at time $k$. Moreover, for the latter case, if Bernoulli component $n_{k|k-1}+j$ has also been associated with any measurements at subsequent time steps, then, due to constraint 6), we cannot change the data association $\theta_k(n_{k|k-1}+j)=j$. That is, in this case we have that $\pi_{k, i}^{\theta_k(i)}=0$. Therefore, we only need to focus on the case that Bernoulli component $n_{k|k-1}+j$ is only detected once at time $k$. 

We denote the data association sequence where each measurement initiates a Bernoulli component as $\hat{\theta}_{1:K}$, which, for each time $k\in\{1,\dots, K\}$, satisfies that $\widehat{\theta}_k(n_{k|k-1}+j)=j$, $j\in\{1,\dots,m_k\}$, and $\widehat{\theta}_t(i) = 0$ for $i \in \{1,\dots,n_{k|k-1}\}$ and $t\in\{k+1,\dots, K\}$. According to the above analysis, the ratio between the factors $\Pi_{k,n_{k|k}}^{\theta_k(i)}$ of the two cases  $\theta_k(i)=j$ (denoted as $\Pi_{k,n_{k|k}}^{\theta_k(i)=j}$), where $j\in\{1,\dots,m_k\}$, and $\theta_k(i)=0$ (denoted as $\Pi_{k,n_{k|k}}^{\theta_k(i)=0}$) is 
\begin{equation}\label{eq_ratio}
  \frac{\Pi_{k,n_{k|k}}^{\theta_k(i)=j}}{\Pi_{k,n_{k|k}}^{\theta_k(i)=0}} = \frac{1}{w_{K|K}^{n_{k|k-1}+j,\widehat{\theta}^{n_{k|k-1}+j}_{1:K}}},
\end{equation}
where $w_{K|K}^{n_{k|k-1}+j,\widehat{\theta}^{n_{k|k-1}+j}_{1:K}}$ is the weight of local hypothesis that Bernoulli component $n_{k|k-1}+j$ is only detected once at time $k$. Dividing \eqref{eq_condition_block} by $\Pi_{k,n_{k|k}}^{\theta_k(i)=0}$ and using \eqref{eq_ratio} yields the following simplified expression of \eqref{eq_condition_block}: 
\begin{equation}
  \pi_{k,i}^{\theta_k(i)}
  \propto 1_{\Theta_{1:K}}(\theta_{1:K})\begin{cases}
    w_{K|K}^{i,\theta_{1:K}^i} & \theta_k(i) = 0 \\
    \frac{w_{K|K}^{i,\theta_{1:K}^i}}{w_{K|K}^{n_{k|k-1}+j,\widehat{\theta}^{n_{k|k-1}+j}_{1:K}}} & \theta_k(i) = j.
  \end{cases}\label{eq_condition_sample2}
\end{equation}
The simplified expression \eqref{eq_condition_sample2} has intuitive interpretations: 1) the conditional probability $\pi_{k,i}^{\theta_k(i)}$ is proportional to the local hypothesis weight $w_{K|K}^{i,\theta_{1:K}^i}$; 2) the conditional probability $\pi_{k,i}^{\theta_k(i)}$ is divided by the local hypothesis weight $w_{K|K}^{n_{k|k-1}+j,\widehat{\theta}^{n_{k|k-1}+j}_{1:K}}$ when $\theta_k(i)=j$ to compensate the case that measurement $z_k^j$ is not associated to new Bernoulli component $n_{k|k-1}+j$ at time $k$.

It is computationally efficient to evaluate \eqref{eq_condition_sample2}, making sampling $\theta_k(i)$ easy. In addition, the weights $w_{K|K}^{n_{k|k-1}+j,\widehat{\theta}^{n_{k|k-1}+j}_{1:K}}$, $j\in\{1,\dots,m_k\}$, $k\in\{1,\dots,K\},$ can be pre-computed to avoid repetitive computations. Starting with any valid $\theta_{1:K}$, the Gibbs sampler draws samples of $(\theta_k(i),\theta_k(n_{k|k-1}+1:n_{k|k}))$ from \eqref{eq_condition_sample2} by iterating  $i\in\{1,\dots,n_{k|k-1}\}$, $k\in\{1,\dots,K\}$. The blocked Gibbs sampler is summarized in Algorithm \ref{alg_gibbs}, and the choice of the initial data associations $\theta^{(0)}$ is described in Section \ref{sec_initial_mc}.

\begin{algorithm}[!ht]
  \small
  \caption{Blocked Gibbs multi-scan data association.}\label{alg_gibbs}
  \begin{algorithmic}[1]
  \REQUIRE Initial data associations $\theta_{1:K}^{(0)}$, number of iterations $T$.
  \ENSURE $\{ \theta_{1:K}^{(\tau)}: \tau = 1,\dots,T \}$.
  \FOR {$\tau = 1$ to $T$}
  \FOR {$k=1$ to $K$}
  \FOR {$i=1$ to $n_{k|k-1}$}
  \STATE Sample $(\theta_k^{(\tau)}(i),\theta_k^{(\tau)}(n_{k|k-1}+1:n_{k|k}))$ via \eqref{eq_condition_sample2}.
  \ENDFOR
  \ENDFOR
  \ENDFOR
  \end{algorithmic}
\end{algorithm}

\textit{Remark: It is possible to move from any $\theta_{1:K}\in \Theta_{1:K}$ to any other $\widetilde{\theta}_{1:K}\in \Theta_{1:K}$ in finite steps. This means that the Markov chain of $\theta_{1:K}$ is irreducible, which is a sufficient condition for the convergence of Gibbs sampler on discrete variables \cite{roberts1994simple}.}

\subsection{Metropolis-Hastings Sampling}\label{sec_mh}

The fast mixing of the Markov chain is essential to obtain an efficient MCMC sampling algorithm \cite{luengo2020survey}. In the above-described method using blocked Gibbs sampling, we change the data associations of an existing Bernoulli component and all new Bernoulli components at the same time step at a time. This, however, may lead to poor mixing of the Markov chain if hypotheses become highly correlated across time and objects, e.g. in scenarios with closely-spaced objects. Moreover, due to constraint 5) of $\theta_{1:K}$, changing the assignment of measurements that have been used to initiate Bernoulli components may take many intermediate steps. A more appealing strategy is to use MH sampling to simultaneously change the measurement association of multiple Bernoulli components across multiple time steps.

To sample from \eqref{eq_full_sample} using MH sampling, we first pick a valid data association ${\theta}_{1:K}^{(0)}$ to initialize the Markov chain. Then we draw a data association ${\theta}_{1:K}$ from a proposal distribution $q(\cdot|{\theta}_{1:K}^{(\tau-1)})$ at the $\tau$-th iteration, which specifies the probability of proposing ${\theta}_{1:K}$ conditioned on ${\theta}_{1:K}^{(\tau-1)}$. The proposed data association ${\theta}_{1:K}$ is accepted with probability 
\begin{equation}\label{eq_accept}
  A\left({\theta}_{1:K},{\theta}_{1:K}^{(\tau-1)}\right) = \min\left(1, \frac{\pi\left({\theta}_{1:K}\right)q\left({\theta}_{1:K}^{(\tau-1)}|{\theta}_{1:K}\right)}{\pi\left({\theta}_{1:K}^{(\tau-1)}\right)q\left({\theta}_{1:K}|{\theta}_{1:K}^{(\tau-1)}\right)} \right).
\end{equation}
If ${\theta}_{1:K}$ is accepted, we set ${\theta}_{1:K}^{(\tau)}={\theta}_{1:K}$, otherwise ${\theta}_{1:K}^{(\tau)}= {\theta}_{1:K}^{(\tau-1)}$. Given enough number of iterations $T$, ${\theta}_{1:K}$ can be regarded as a sample from the stationary distribution \eqref{eq_full_sample}. 

\begin{figure}[!t]
\centering
\includegraphics[width=\linewidth]{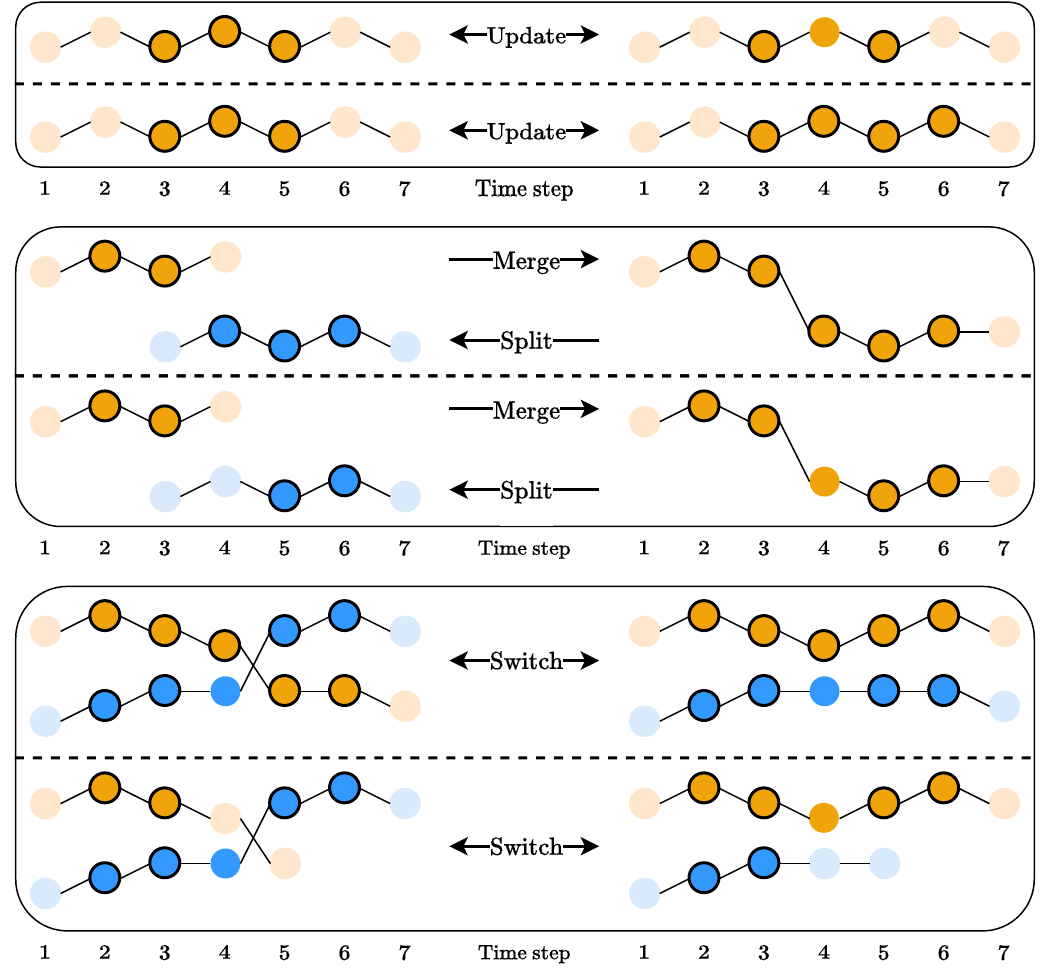}
\caption{Examples of four proposal moves. Each proposal move is illustrated with two examples, separated using a dash line. The object state sequences are represented using chained circles. Circles with/without borderline represent measurement detection/misdetection. In addition, circles in light orange/blue represent object states with existence uncertainty. For instance, the trajectory in the top left corner has a maximum length of 7. It is first detected at time 3 and last detected at time 5, but it may start at time 1, 2, or 3 and end at time 5, 6, or 7.}
\label{fig_proposalMove}
\end{figure}

To develop efficient MH sampling algorithms, it is important to construct proposal distributions $q(\cdot|\cdot)$ that can well explore the data association space and are straightforward to evaluate. The proposal distributions for batch TPMBM implementation using MH sampling consist of four moves: 1) track update move, 2) merge/split move, and 3) track switch move. An illustration of these moves is given in Fig. \ref{fig_proposalMove}. For convenience, we index each type of move by $c\in\{1,2,3,4\}$, with $c=1$ for the track update move, $c=2$ for the merge move, $c=3$ for the split move, and finally $c=4$ for the track switch move.

Each move is chosen randomly from a category distribution $p(c)$, which should be constructed according to the MB density conditioned on ${\theta}_{1:K}$ and the tracking scenario of interest. For instance, in challenging MOT scenarios with closely-spaced objects, we could increase the probability of performing track switch move $p(c=4)$. The proposal distribution $q(\cdot|\cdot)$ is constructed from a mixture of the form \cite[Eq. (11.42)]{Bishop:2006}
\begin{equation}\label{eq_proposal}
  q\left(\widetilde{\theta}_{1:K}|\theta_{1:K}\right) = \sum_{c=1}^4p(c)q_c\left(\widetilde{\theta}_{1:K}|\theta_{1:K}\right),
\end{equation}
where auxiliary variable $c$ determines the distribution $q_c(\cdot|\cdot)$ to be sampled and $q_{c}(\widetilde{\theta}_{1:K}|\cdot)$ takes non-zero values only if $\widetilde{\theta}_{1:K}$ is obtained by taking the corresponding move $c$. 

We proceed to describe each proposal move conditioned on a valid data association $\theta_{1:K}$ in detail.

\subsubsection{Track update move}\label{sec_track_update_move}
We first sample a Bernoulli component $i$ from the index set of Bernoulli components that are associated with at least two measurements uniformly at random (u.a.r.) 
\begin{equation}\label{eq_ber_r1}
  \mathbb{I}_{r=1} = \left\{ i\in \{1,\dots,n_{K|K}\}: r_{K|K}^{i,\theta^i_{1:K}} = 1 \right\},
\end{equation}
where $r_{K|K}^{i,\theta^i_{1:K}}$ is the existence probability of the $\theta_{1:K}^i$-th local hypothesis under the $i$-th Bernoulli component that is included in global hypothesis $\Lambda_K(\theta_{1:K})$.

For the selected Bernoulli component $i$, we denote its first detected time step as 
\begin{equation}
  t_{\text{first}}^i = \min \left( \left\{k\in\{1,\dots,K\}: \exists \theta_k(i) > 0 \right\} \right),
\end{equation}
and we denote its maximum end time as 
\begin{equation}
  t_{\text{end}}^i = \max \left( \left\{e_{K|K}^{i,\theta_{1:K}^i,l}\right\}_{l=1}^{L_{K|K}^{i,\theta_{1:K}^i}}  \right),
\end{equation}
where $e_{K|K}^{i,\theta_{1:K}^i,l}$ represents the trajectory end time of the $l$-th component in the trajectory density of the $\theta_{1:K}^i$-th local hypothesis of the $i$-th Bernoulli component. Then we sample a time $t$ u.a.r. from the time interval $t_{\text{first}}^i+1:t_{\text{end}}^i$ and draw a random sample of $({\theta}_t(i),{\theta}_t(n_{t|t-1}+1:n_{t|t}))$ from \eqref{eq_condition_sample2} with $k=t$ via blocked Gibbs sampling described in Section \ref{sec_gibbs}.

Under the above setting, the proposal distribution is 
\begin{equation}
  q_1\left(\widetilde{\theta}_{1:K}|{\theta}_{1:K}\right) = \frac{\pi_{t,i}^{\widetilde{\theta}_t(i)}}{|\mathbb{I}_{r=1}|\left(t_{\text{end}}^i-t_{\text{first}}^i\right)}.
\end{equation}
The Gibbs sampling procedure can be regarded as a particular instance of the MH algorithm where the acceptance probability is one \cite[Eq. (11.49)]{Bishop:2006}. Thus, the MH sampling steps for track update moves are always accepted.

\textit{Remark: When $p(c=1) = 1$, the presented MH sampler reduces to a Gibbs sampler, where variables $({\theta}_t(i),{\theta}_t(n_{t|t-1}+1:n_{t|t}))$ to be updated at each step are chosen at random instead of by cycling $i,t$ in some particular order\footnote{In Gibbs sampling, two common scan orders for sampling variables are random scan and sequential scan. In practice, sequential scans are often used since they are superior from the hardware efficiency perspective \cite{smola2010architecture}.}.}

\subsubsection{Merge/split move}

For a merge move, we first sample a Bernoulli component $i$ u.a.r. from the index set of Bernoulli components that have been assigned with at least one measurement
\begin{equation}\label{eq_ber_r0}
  \mathbb{I}_{r>0} = \left\{ i\in \{1,\dots,n_{K|K}\}: r_{K|K}^{i,\theta^i_{1:K}} > 0 \right\}.
\end{equation}
For the selected Bernoulli component $i$, we denote its last detected time step as 
\begin{equation}
  t_{\text{last}}^i = \max \left( \left\{k\in\{1,\dots,K\}: \exists \theta_k(i) > 0 \right\} \right).
\end{equation}
Then we sample a trajectory Bernoulli component $i^\prime$ u.a.r. from the index set of Bernoulli components that are either initiated after time $t_{\text{last}}^i$ or last detected before time $t_{\text{first}}^i$
\begin{equation}
  \mathbb{I}_{t^i} = \left\{ i^\prime\in\{1,\dots,n_{K|K}\}: t^{i^\prime}_{\text{first}} > t_{\text{last}}^i ~\text{or}~ t^{i^\prime}_{\text{last}} < t_{\text{first}}^i \right\}.
\end{equation} 
In a merge move, we first let $\widetilde{\theta}_{1:K} = \theta_{1:K}$, and then
\begin{itemize}
  \item If $t^{i^\prime}_{\text{first}} > t_{\text{last}}^i$, we set $\widetilde{\theta}_k(i) = \theta_k(i^\prime)$ and $\widetilde{\theta}_k(i^\prime) = 0$ for $k \in \{t^{i^\prime}_{\text{first}}, \dots, t^{i^\prime}_{\text{last}}\}$.
  \item If $t^{i^\prime}_{\text{last}} < t_{\text{first}}^i$, we set $\widetilde{\theta}_k(i^\prime) = \theta_k(i)$ and $\widetilde{\theta}_k(i) = 0$ for $k \in \{t^{i}_{\text{first}}, \dots, t^{i}_{\text{last}}\}$.
\end{itemize}
After merging, one of the two Bernoulli components is not associated with any measurements, and this Bernoulli component has local hypothesis weight one and probability of existence zero (cf. \eqref{eq_nonexist}). 

For a split move, we first sample a Bernoulli component $i$ u.a.r. from $\mathbb{I}_{r=1}$ \eqref{eq_ber_r1}. Then we sample a time $t$ u.a.r. from the set of time steps (except for $t^{i}_{\text{first}}$) that Bernoulli component $i$ is detected
\begin{equation}
  \mathbb{I}_{\theta_k(i)} = \left\{k\in\{t^{i}_{\text{first}}+1,\dots,t^{i}_{\text{last}}\}: \exists \theta_k(i) > 0 \right\}.
\end{equation}
In a split move, we first let $\widetilde{\theta}_{1:K} = \theta_{1:K}$, and then we set $\theta_k(i) = 0$ and $\widetilde{\theta}_k(i^\prime) = \theta_k(i)$ where $i^\prime = \sum_{k^\prime=1}^{t-1}m_{k^\prime} + \theta_t(i)$ for $k\in\{t,\dots,t^{i}_{\text{last}}\}$. That is, after splitting, Bernoulli component $i$ is not associated with any measurements since time $t$, and Bernoulli component $i^\prime$ initiated by measurement $z_t^{\theta_t(i)}$ at time $t$ inherits the measurement association of Bernoulli component $i$ after time $t$.

The probability of proposing a merge move $q_2(\cdot|\cdot)$ or a split move $q_3(\cdot|\cdot)$ is easy to compute since both moves are sampled u.a.r. Specifically, we have 
\begin{align}
  q_2\left(\widetilde{\theta}_{1:K}|{\theta}_{1:K}\right) &= \frac{1}{|\mathbb{I}_{r>0}||\mathbb{I}_{t^i}|},\label{eq_merge_proposal}\\
  q_3\left(\widetilde{\theta}_{1:K}|{\theta}_{1:K}\right) &= \frac{1}{|\mathbb{I}_{r=1}||\mathbb{I}_{\theta_k(i)}|} \label{eq_split_proposal}. 
\end{align}
To evaluate the acceptance probability \eqref{eq_accept}, we also need to compute the ratio of the two data association probabilities $\pi(\cdot)$ after and before the move. For the merge move, this ratio is 
\begin{equation}
  \frac{\pi\left(\widetilde{\theta}_{1:K}\right)}{\pi(\theta_{1:K})} = \begin{cases}
    \frac{w_{K|K}^{i,\widetilde{\theta}_{1:K}^i}}{w_{K|K}^{i,\theta_{1:K}^i}w_{K|K}^{i^\prime,\theta_{1:K}^{i^\prime}}} & t^{i^\prime}_{\text{first}} > t_{\text{last}}^i \\
    \frac{w_{K|K}^{i^\prime,\widetilde{\theta}_{1:K}^{i^\prime}}}{w_{K|K}^{i,\theta_{1:K}^i}w_{K|K}^{i^\prime,\theta_{1:K}^{i^\prime}}} & t^{i^\prime}_{\text{last}} < t_{\text{first}}^i.
  \end{cases}\label{eq_merge}
\end{equation} 
As for the split move, the ratio is given by 
\begin{equation}
  \frac{\pi\left(\widetilde{\theta}_{1:K}\right)}{\pi(\theta_{1:K})} = \frac{w_{K|K}^{i,\widetilde{\theta}_{1:K}^i}w_{K|K}^{i^\prime,\widetilde{\theta}_{1:K}^{i^\prime}}}{w_{K|K}^{i,{\theta}_{1:K}^i}}.\label{eq_split}
\end{equation}

\subsubsection{Track switch move}\label{sec_track_switch_move}
We first sample two Bernoulli components $i$, $i^\prime$ ($i\neq i^\prime$) u.a.r. from $\mathbb{I}_{r=1}$ \eqref{eq_ber_r1}. Then we sample a time $t$ u.a.r. from the time interval
\begin{equation*}
  \max\left(t^{i}_{\text{first}},t^{i^\prime}_{\text{first}}\right)+1:\max\left(t^{i}_{\text{last}},t^{i^\prime}_{\text{last}}\right).
\end{equation*}
To perform a track switch move, we first let $\widetilde{\theta}_{1:K} = \theta_{1:K}$, and then we swap the measurement associations $\widetilde{\theta}_k(i)$ and $\widetilde{\theta}_k(i^\prime)$ for $k\in\{t,\dots,\max(t^{i}_{\text{last}},t^{i^\prime}_{\text{last}})\}$. Since the track switch move is self-reversible, it holds that $q_4(\widetilde{\theta}_{1:K}|\theta_{1:K}) = q_4(\theta_{1:K}|\widetilde{\theta}_{1:K})$. The ratio of the data association weights after and before the move is 
\begin{equation}
  \frac{\pi\left(\widetilde{\theta}_{1:K}\right)}{\pi(\theta_{1:K})} = \frac{w_{K|K}^{i,\widetilde{\theta}_{1:K}^i}w_{K|K}^{i^\prime,\widetilde{\theta}_{1:K}^{i^\prime}}}{w_{K|K}^{i,{\theta}_{1:K}^i}w_{K|K}^{i^\prime,{\theta}_{1:K}^{i^\prime}}}.\label{eq_switch}
\end{equation}

The choice of these four moves can be justified as follows. For changing the measurement association history of a single Bernoulli component, we may either change the measurement association at a single time step, or split the measurement association history into two parts. For simultaneously changing the measurement association history of two Bernoulli components, we may swap the measurement associations at a single time step, or merge the two measurement association histories.
The MH sampler is summarized in Algorithm \ref{alg_mh}. 

\begin{algorithm}[!ht]
  \small
  \caption{Metropolis-Hastings multi-scan data association.}\label{alg_mh}
  \begin{algorithmic}[1]
  \REQUIRE Initial data associations $\theta_{1:K}^{(0)}$, number of iterations $T$.
  \ENSURE $\{ \theta_{1:K}^{(\tau)}: \tau = 1,\dots,T \}$.
  \FOR {$\tau = 1$ to $T$}
  \STATE Sample $c \sim p(c)$.
  \IF{$c==1$ (track update move)}
    \STATE Sample $\theta_{1:K}\sim q_1(\widetilde{\theta}_{1:K}|\theta_{1:K})$.
    \STATE $A({\theta}_{1:K},{\theta}_{1:K}^{(\tau-1)}) = 1$.
  \ELSIF{$c==2$ (merge move)}
    \STATE Sample $\theta_{1:K}\sim q_2(\widetilde{\theta}_{1:K}|\theta_{1:K})$.
    \STATE Compute $A({\theta}_{1:K},{\theta}_{1:K}^{(\tau-1)})$ using \eqref{eq_merge_proposal}, \eqref{eq_split_proposal} and \eqref{eq_merge}.
  \ELSIF{$c==3$ (split move)}
  \STATE Sample $\theta_{1:K}\sim q_3(\widetilde{\theta}_{1:K}|\theta_{1:K})$.
  \STATE Compute $A({\theta}_{1:K},{\theta}_{1:K}^{(\tau-1)})$ using \eqref{eq_merge_proposal}, \eqref{eq_split_proposal} and \eqref{eq_split}.
  \ELSIF{$c==4$ (track switch move)}
  \STATE Sample $\theta_{1:K}\sim q_4(\widetilde{\theta}_{1:K}|\theta_{1:K})$.
  \STATE Compute $A({\theta}_{1:K},{\theta}_{1:K}^{(\tau-1)})$ using \eqref{eq_switch}.
  \ENDIF
  \STATE Sample $u$ u.a.r. from $[0,1]$.
  \IF{$u < A({\theta}_{1:K},{\theta}_{1:K}^{(\tau-1)})$}
    \STATE $\theta_{1:K}^{(\tau)} = {\theta}_{1:K}$.
  \ELSE
    \STATE $\theta_{1:K}^{(\tau)} = \theta_{1:K}^{(\tau-1)}$.
  \ENDIF
  \ENDFOR
  \end{algorithmic}
\end{algorithm}

Several key distinctions exist between the proposal distribution presented here and the proposal distributions in \cite{oh2009markov}. First, there are no birth/death and extension/reduction moves, since uncertainties on trajectory existence and start/end times are captured via single trajectory densities. That is, these events are never sampled in TPMBM; instead they are analytically marginalized out given a global hypothesis. Note that this is conceptually analogous to a Rao-Blackwellized particle filter \cite{schon2005marginalized}, where some elements in the state are marginalized out analytically to obtain a more effective algorithm. Second, the proposed track update moves using Gibbs sampling are always accepted, allowing faster convergence of the Markov chain. Third, the track switch move presented here considers more cases, which enables the swap of measurement association sequence and consecutive misdetections, see the illustration at the bottom of Fig. \ref{fig_proposalMove} (below the dash line) for an example. This is not possible for the track switch move in the MCMC implementations presented in \cite{oh2009markov,vu2014particle}, which only covers the case of swapping measurement associations, as illustrated by the other example at the bottom of Fig. 3 (above the dash line). Lastly, the data association space does not need to be constrained by incorporating domain-specific knowledge, such as maximum object speed or minimum track length. 

\textit{Remark: To show that the presented MH sampler converges, it is sufficient to show that the proposal distribution \eqref{eq_proposal} is irreducible and aperiodic \cite{roberts1994simple}. Since it is possible to move from any $\theta_{1:K}$ to any other $\widetilde{\theta}_{1:K}$ in finite steps via the track update moves, the proposal distribution \eqref{eq_proposal} is irreducible. In addition, it is trivial to see that there is always a positive probability that $\theta_{1:K}^{(\tau)} = \theta_{1:K}^{(\tau-1)}$ (in the track update moves), so the proposal distribution \eqref{eq_proposal} is also aperiodic.}

\subsection{Practical Considerations}
\subsubsection{Initialization of the Markov Chain}\label{sec_initial_mc}
The above-described MCMC sampling methods can be initialized with any valid data associations $\theta_{1:K}$. A simple initialization choice is by setting $\theta_k(n_{k|k-1}+j) = j$ for $j\in\{1,\dots,m_k\}$ and all other $\theta_k(i)$ to zero for $k\in\{1,\dots, K\}$, which corresponds to the case that each measurement initiates a new Bernoulli component. Better initial data associations $\theta_{1:K}^{(0)}$ can be obtained by running a PMBM filter or a PMB filter with the global nearest neighbour data association. 

\subsubsection{Approximations}
The presented MCMC sampling methods can be made more computationally efficient by 1) reducing the data association space, and 2) simplifying the computation of local hypothesis densities and weights.

First, we use gating to remove unlikely measurement associations. Second, for local hypotheses corresponding to newly detected trajectories at time $k$, we set $r^{i,a^i}_{k|k} = 0$ if $r^{i,a^i}_{k|k}$ is less than a threshold; these local hypotheses do not need to be predicted or updated in subsequent time steps. Third, we prune mixture components in the trajectory density of local hypotheses with small weights, which amounts to truncating the pmf of trajectory start and end times. After pruning, if the probability that a previously detected Bernoulli component is alive reduces to zero, it remains unchanged in subsequent time steps. We also prune mixture components in the PPP intensity of undetected trajectories with small weights.

In addition, to compute the weights of local hypotheses after changing the data associations in MCMC sampling, we need to run the Bernoulli prediction and update equations for the new data association sequence. Instead of computing the smoothed object state densities in each sampling iteration, we use $L=1$-scan approximation \cite{garcia2019trajectory}. Under this setting, single object states at individual time steps are considered independent and only the latest object state of alive trajectories is updated.

\subsubsection{Estimation}
From the MCMC sampling results, we find unique data association samples and their corresponding global hypotheses. Since all distinct global hypotheses will reduce the $L_1$ norm error in truncation \cite[Sec. V-D]{granstrom2019poisson}, burn-in time and correlation between consecutive samples are not relevant when approximating the TPMBM posterior. To report a reasonable set of trajectories estimate, we first identify a global hypothesis using, e.g. any of the three estimators described in \cite{garcia2018poisson}. Then for each local hypothesis, we extract the maximum a posteriori trajectory start/end time estimates. Finally, we apply backward smoothing to the object state filtering densities to obtain smoothed trajectory estimates. 

\section{Simulations and Results}\label{sec_results}

We compare the multi-trajectory estimation performance of the following multi-object trackers with their linear Gaussian implementations:
\begin{enumerate}
  \item TPMBM filter using Murty's algorithm \cite{granstrom2018poisson}, referred to as TPMBM-M.
  \item TPMBM filter using dual decomposition \cite{xia2019multi}, referred to as TPMBM-DD.
  \item Batch TPMBM filter using Gibbs sampling, referred to as B-TPMBM-G.
  \item Batch TPMBM filter using MH sampling, referred to as B-TPMBM-MH\footnote{MATLAB code is available at \url{https://github.com/yuhsuansia}.}.
  \item Labelled trajectory $\text{MBM}_{01}$ filter with partial smoothing \cite{nguyen2019glmb} using Murty's algorithm, referred to as T$\text{MBM}_{01}$-M.
  \item Batch labelled trajectory $\text{MBM}_{01}$ filter\footnote{This is also called multi-scan generalized labelled MB filter.} using multi-scan Gibbs sampling \cite{vo2019multi}, referred to as B-T$\text{MBM}_{01}$-G.
\end{enumerate}

\begin{figure}[!t]
  \centering
  \includegraphics[width=0.95\linewidth]{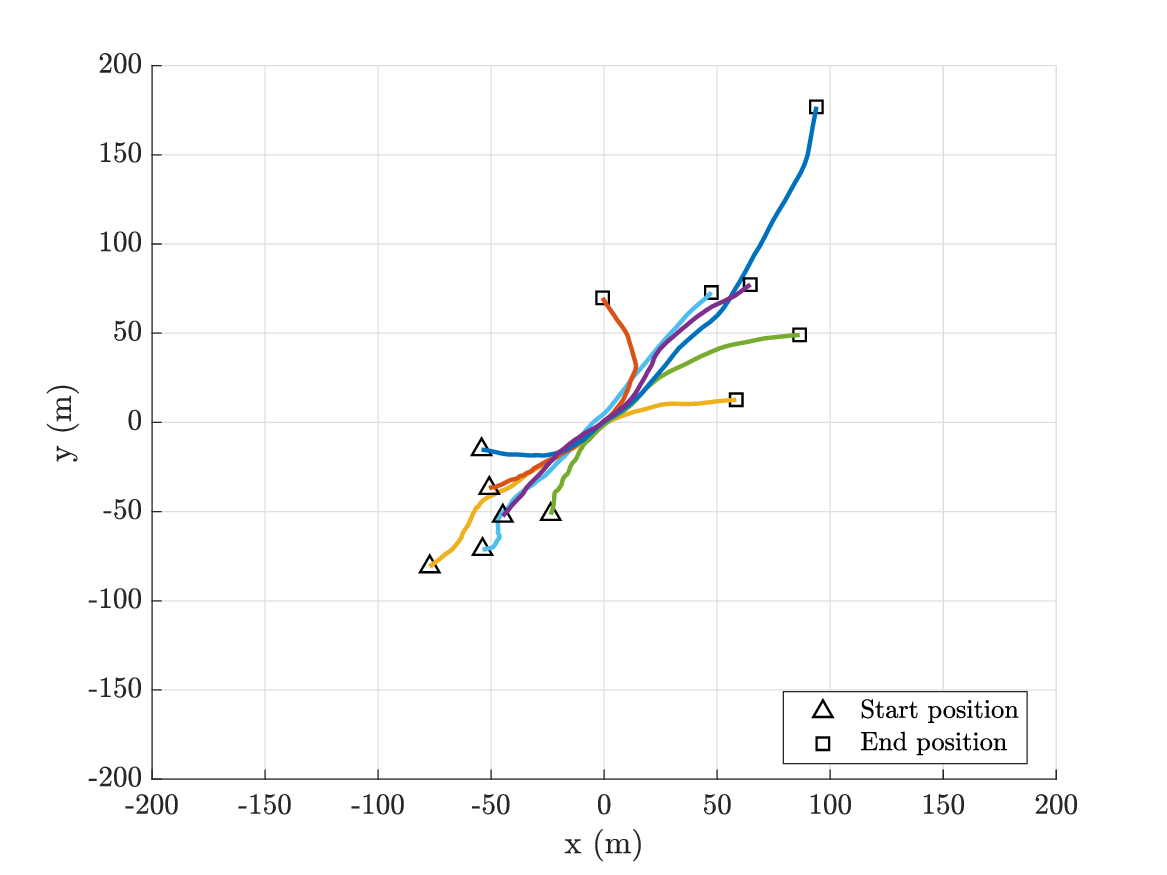}
  \caption{Ground truth of the simulated scenario, which give rise to a difficult data association problem. The start/end positions of trajectories are marked by triangles/squares. Six objects are born at time 1, 1, 11, 11, 21, and 21, and die at time 61, 61, 71, 71, 81, and 81, respectively. }
  \label{fig_groundTruth}
\end{figure}

Both TPMBM-M and TPMBM-DD are implemented in the Gaussian information form \cite{granstrom2018poisson} that enables smoothing-while-filtering. For all the other implementations, we apply Rauch-Tung-Striebel smoothing to the object state filtering densities conditioned on the mostly likely global hypothesis at the final time step to extract smoothed trajectory estimates.

\begin{figure*}[!t]
  \centering
  \includegraphics[width=0.3\linewidth]{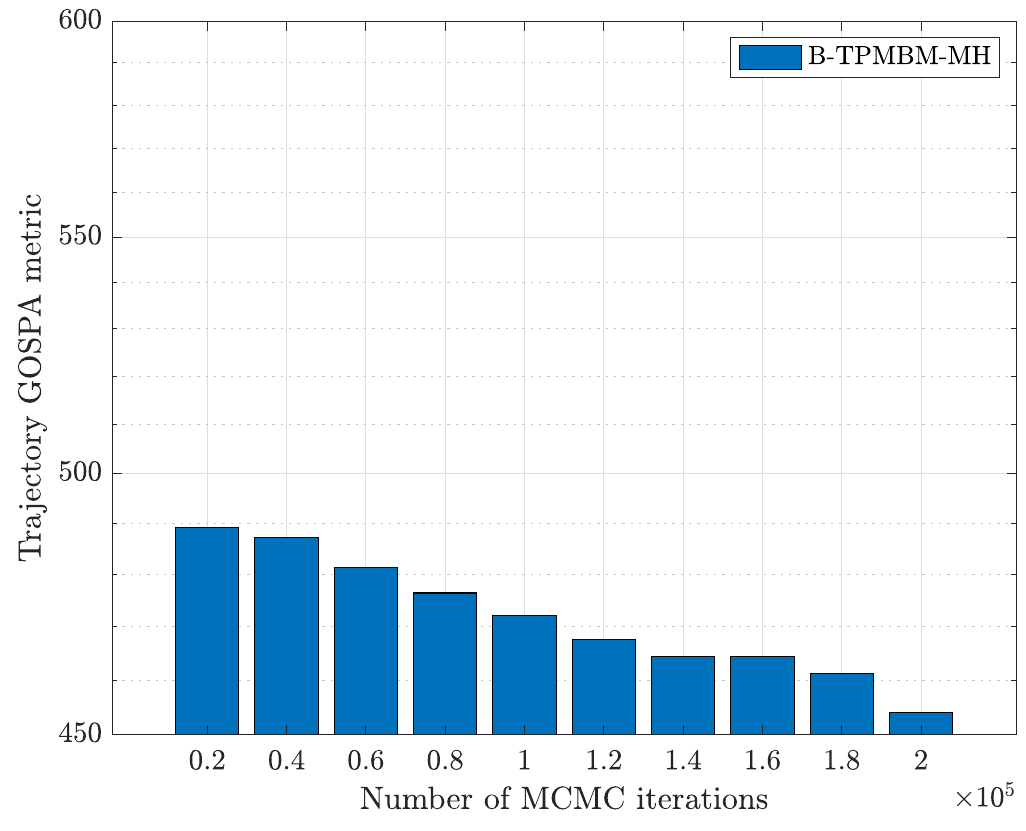}
  \includegraphics[width=0.3\linewidth]{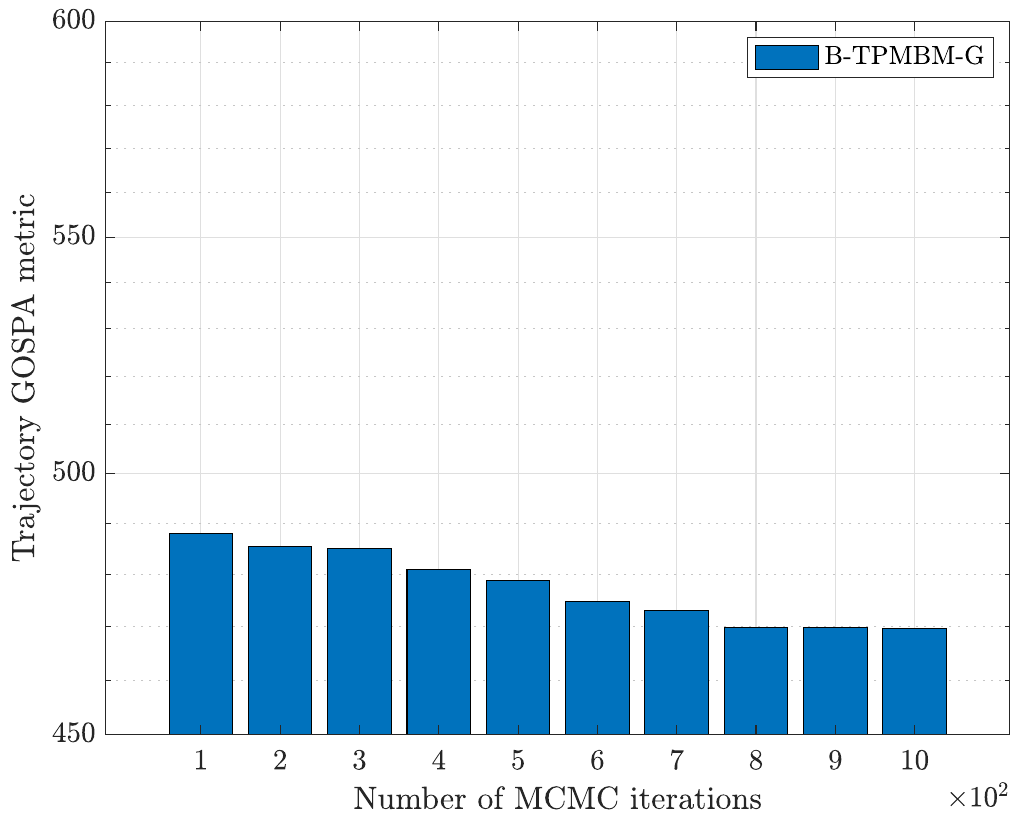}
  \includegraphics[width=0.3\linewidth]{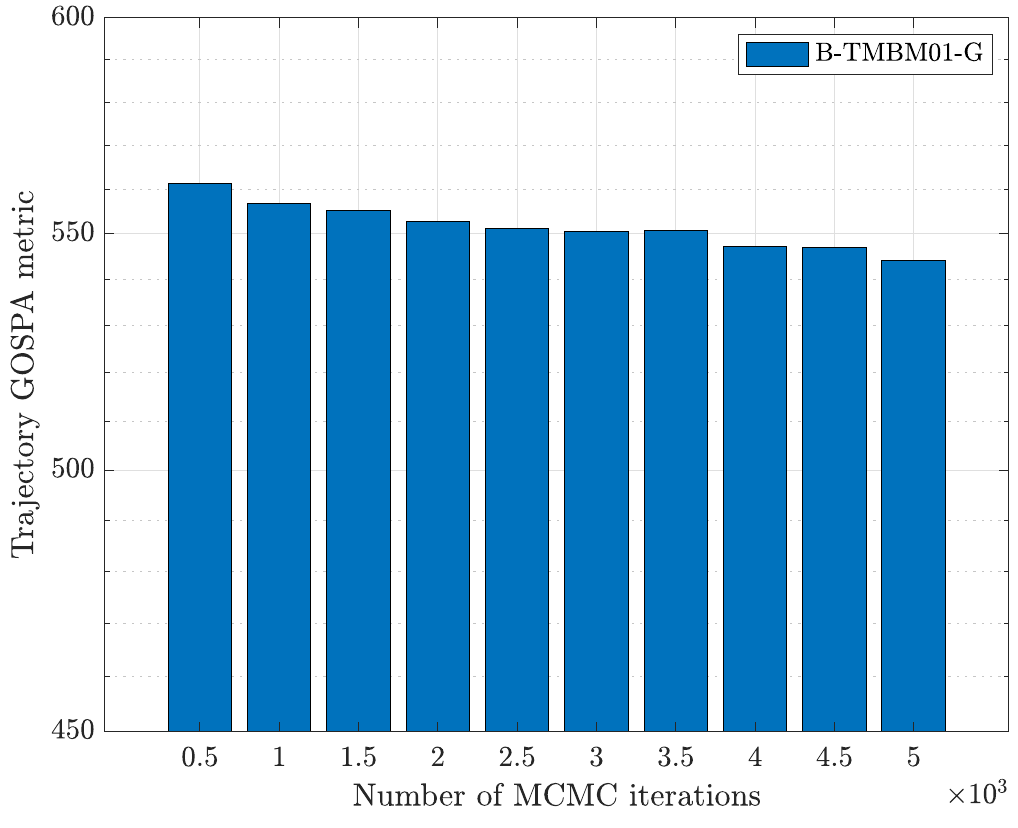}
  \caption{Trajectory GOSPA metric error versus the number of MCMC sampling iterations for B-TPMBM-MH, B-TPMBM-G, and $\text{B-TMBM}_{01}$. Note that the $x$-axis in the three figures has different scales.}
  \label{fig_MCMCiteration}
\end{figure*}

In simple scenarios where objects are well spaced, or only a few objects are closely spaced for a limited time steps, the data association uncertainty is low and different implementations present similar tracking performance. For more complex scenarios with higher data association uncertainty, the tracking performance difference among different implementations start to reveal, and it would increase with the number of objects. To this end, we examine a two-dimensional scenario consisting of 81-time steps, where six objects, initially separate, undergo motion that brings them into close proximity before they separate again, in the area $[-200~\text{m}, 200~\text{m}] \times [-200~\text{m}, 200~\text{m}]$. The ground truth of the simulated scenario is illustrated in Fig. \ref{fig_groundTruth}. Each object survives with probability $p^S=0.98$, and each survived object is moving according to a nearly constant velocity motion model $g(x_{k+1}|x_k) = \mathcal{N}(Fx_{k},Q)$ with
\begin{equation*}
  F = I_2\otimes\begin{bmatrix}
    1 & T_s \\ 0 & 1
  \end{bmatrix},\quad Q = 0.09I_2\otimes\begin{bmatrix}
    T_s^3/2 & T_s^2/2 \\ T_s^2/2 & T_s
  \end{bmatrix},
\end{equation*}
where $F$ is the state transition matrix, $Q$ is the motion noise covariance, $T_s = 1~\text{s}$ is the sampling period, and $I_2$ is an identity matrix. 

For TPMBM implementations, the Poisson birth intensity is a Gaussian mixture with six components, each covering a birth location. For each Gaussian component, its mean is given by its corresponding true object birth state rounded to the nearest integer, with weight $0.01$ and covariance $4I_2$. For T$\text{MBM}_{01}$-M and B-T$\text{MBM}_{01}$-G, the MB birth also contains six Bernoulli components, with the same probability hypothesis density as the PPP birth.

Each object is detected with probability $p^D = 0.7$, and, if detected, it generates a measurement according to a linear and Gaussian measurement model $\ell(z|x) = \mathcal{N}(Hx,R)$ with
\begin{equation*}
  H = I_2\otimes\begin{bmatrix}
    1 & 0
  \end{bmatrix},\quad R = I_2,
\end{equation*}
where $H$ is the observation matrix and $R$ is the measurement noise covariance. The clutter is uniformly distributed in the area of interest with Poisson rate $\lambda^C = 30$. 

All the implementations use ellipsoidal gating with a gating size of probability 0.999. For TPMBM implementations, we remove Gaussian components in the Poisson intensity with weight smaller than $10^{-4}$. In addition, we truncate the pmfs of the trajectory birth and end times with probability smaller than $10^{-2}$ and $10^{-4}$, respectively. For online implementations TPMBM-M and TPMBM-DD, we also prune Bernoulli components with existence probability smaller than $10^{-4}$. For both TPMBM-M and T$\text{MBM}_{01}$-M, the maximum number of global hypotheses is $10^4$, and we also prune global hypotheses with weight smaller than $10^{-5}$. For TPMBM-DD, the sliding window has length $7$.

For batch TPMBM, the Markov chain is initialized using the data association, extracted from the global hypothesis with the highest weight by running a TPMBM-M with maximum number of global hypotheses $10^3$ and weight pruning threshold $10^{-4}$. In B-T$\text{MBM}_{01}$-G, the Markov chain is initialized using T$\text{MBM}_{01}$-M with the same parameter setting. For all the implementations, trajectory estimates are obtained from Bernoulli components with existence probability $r=1$ included in the global hypothesis (MB component) with the highest weight. 

We evaluate the multiple trajectory estimation performance using the trajectory generalized optimal sub-pattern assignment (GOSPA) metric\footnote{A concise introduction of the trajectory GOSPA metric can be also found at Appendix E of \url{https://arxiv.org/pdf/2207.10164.pdf}.} \cite{garcia2020metric} parameterized by order 1, cut-off distance 10, and track switch penalty 2. The trajectory GOSPA metric can be broken down into components corresponding to errors in state estimation errors, missed detection errors, false detection errors, and track switch errors.

All the simulation results are obtained through 100 Monte Carlo runs. For the proposal distributions used in B-TPMBM-MH, we have tried different settings, all with $2\times10^5$ MCMC sampling iterations, and found that the setting with high track switching probability yielded the best performance, see Table \ref{tab_ablation} for an ablation study. This observation is aligned with the intuition that in scenarios with objects moving close to each other, for fast mixing of the Markov chain, it is more efficient to swap the associations of Bernoulli components instead of updating the association of one Bernoulli component at a time. In addition, it is computationally more efficient to perform a track switch move than a track update move.

The total number of sampling iterations used in B-TPMBM-MH, B-TPMBM-G and B-T$\text{MBM}_{01}$-G are $2\times10^5$, $10^3$ and $5\times10^3$, respectively. Under this setting, the average runtime for these three implementations is $769~\text{s}$, $1241~\text{s}$ and $1354~\text{s}$, respectively. 
Fig. \ref{fig_MCMCiteration} shows how the average trajectory GOSPA error asymptotically decreases with the number of sampling iterations. Together with the average runtime, it is not difficult to see that B-TPMBM-G consistently outperforms $\text{B-TMBM}_{01}$-G in terms of runtime and trajectory GOSPA. It should be also noted that B-TPMBM-MH performs on par with B-TPMBM-G using about half of its total sampling iterations (and about one-fourth of its runtime). 

\begin{figure}[!t]
  \centering
  \includegraphics[width=0.95\linewidth]{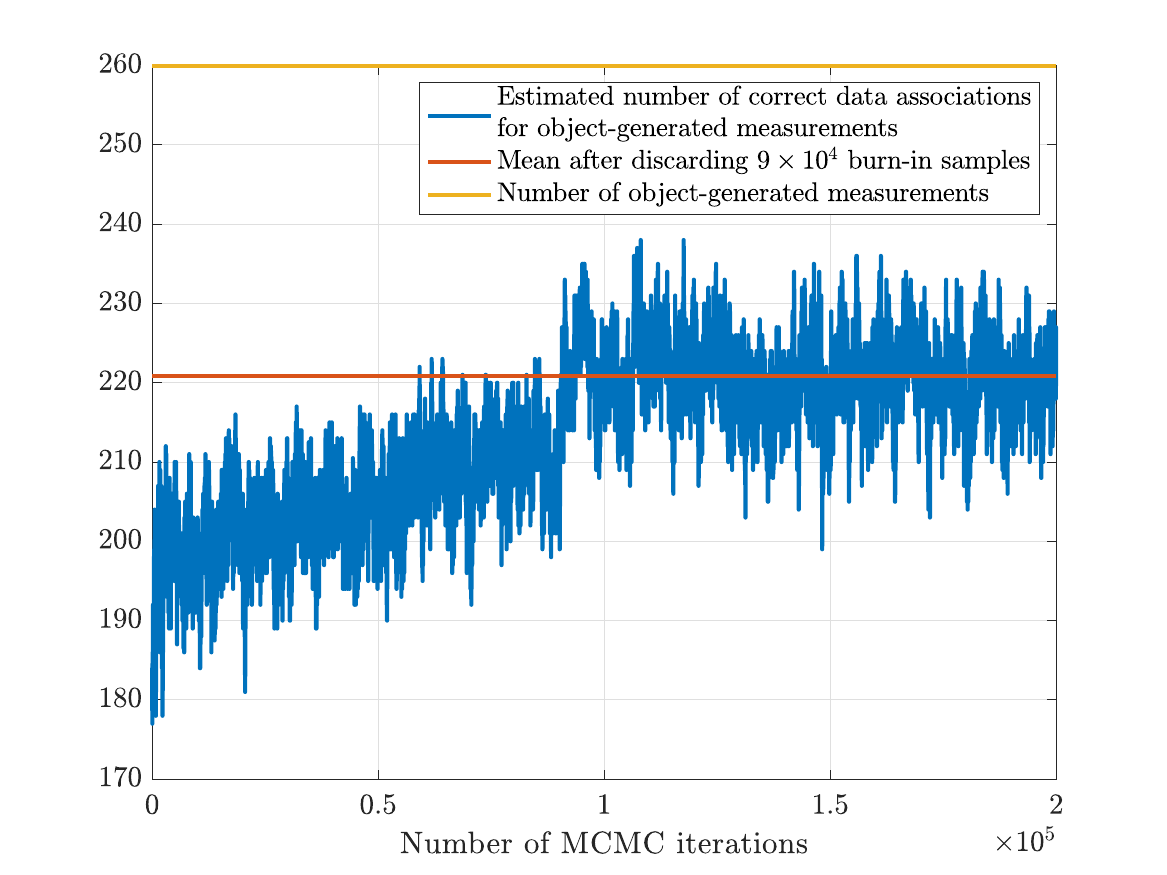}
  \caption{Trace plot of number of correct data associations for object-generated measurements for TPMBM with Gibbs sampling. The quantity of interest is computed every 20 sampling iterations with $2\times10^5$ iterations in total.}
  \label{fig_gibbs_trace}
\end{figure}

\begin{figure}[!t]
  \centering
  \includegraphics[width=0.95\linewidth]{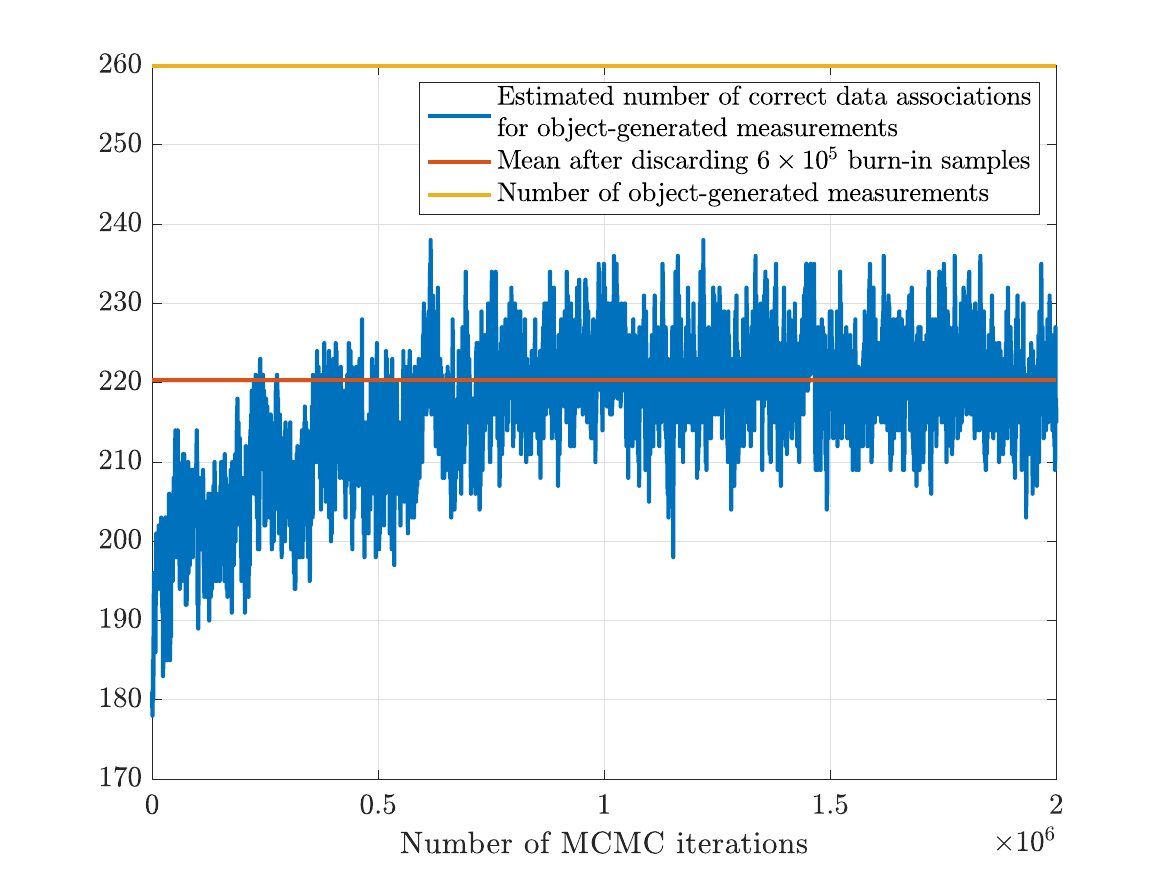}
  \caption{Trace plot of number of correct data associations for object-generated measurements for TPMBM with MH sampling. The quantity of interest is computed every 200 sampling iterations with $2\times10^6$ iterations in total.}
  \label{fig_mh_trace}
\end{figure}

To study the convergence property of MCMC sampling in B-TPMBM-G and B-TPMBM-MH, we present the trace plots of the number of correct data associations for object-generated measurements for one Monte Carlo run. This quantity is computed by leveraging the TGOSPA metric \cite{garcia2020metric}. Specifically, for each pair of matched true object state and estimated object state, we check if the measurement generated by the true object state is associated to the estimated object state. The trace plots for B-TPMBM-G and B-TPMBM-MH are presented in Fig. \ref{fig_gibbs_trace} and Fig. \ref{fig_mh_trace}, respectively. It can be seen from the results that the number of burn-in samples in B-TPMBM-G and B-TPMBM-MH is, respectively, about $9\times10^4$ and $6\times10^5$. Considering that the average runtime of a single sampling iteration in B-TPMBM-MH is only about $1/300$ of it in B-TPMBM-G, B-TPMBM-MH converges to the stationary distribution much faster than B-TPMBM-G. This observation aligns with the results on the TGOSPA error over the number of sampling iterations in Fig. \ref{fig_MCMCiteration}. In addition, when the Markov chain has reached its stationary distribution, the mean of the number of correct data associations in B-TPMBM-G and B-TPMBM-MH is almost the same. This indicates that the stationary distributions obtained by B-TPMBM-G and B-TPMBM-MH may be approximately the same.

\begin{figure}[!t]
  \centering
  \includegraphics[width=0.95\linewidth]{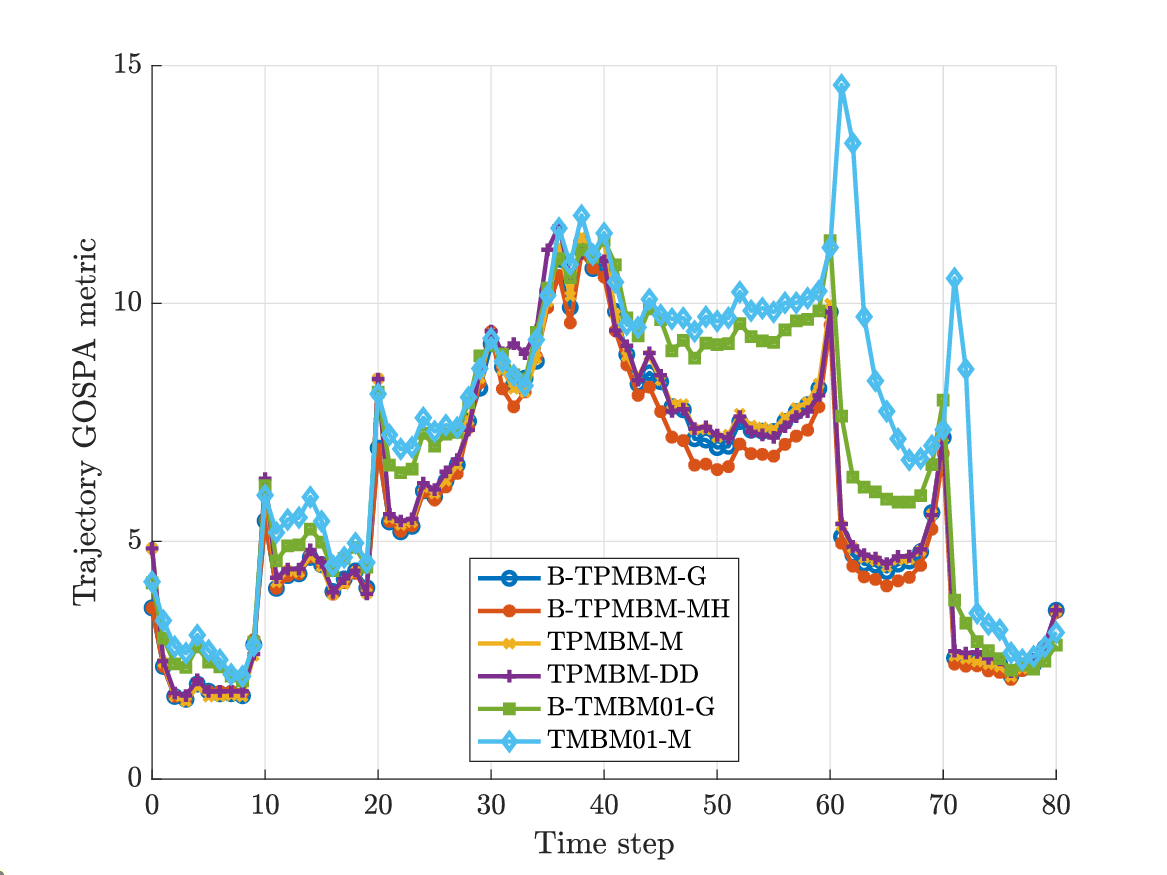}
  \caption{Trajectory GOSPA errors versus time.}
  \label{fig_LPmetric}
\end{figure}

\begin{figure*}[!t]
  \centering
  \includegraphics[width=0.24\linewidth]{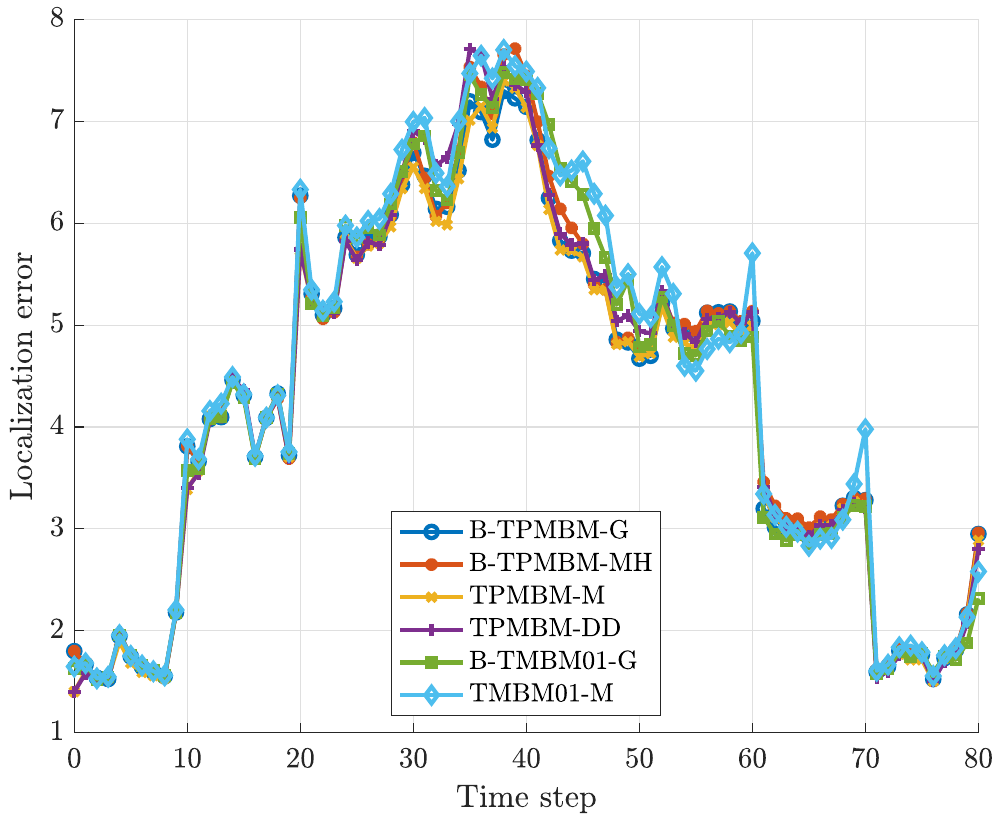}
  \includegraphics[width=0.24\linewidth]{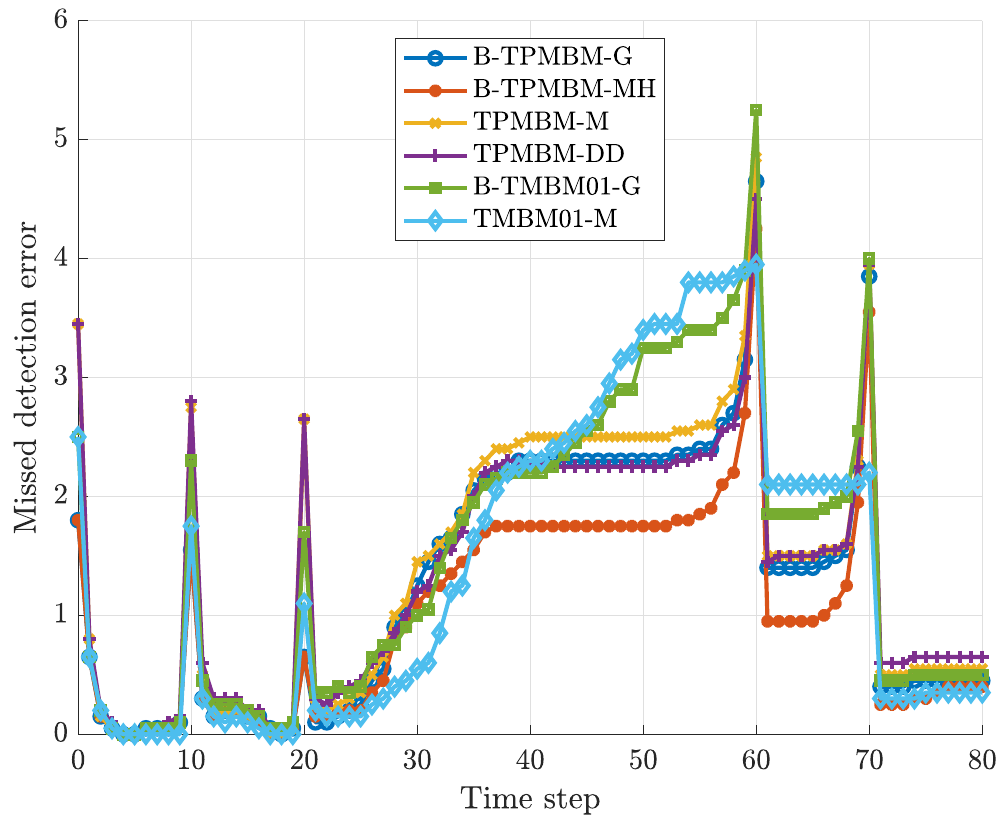}
  \includegraphics[width=0.24\linewidth]{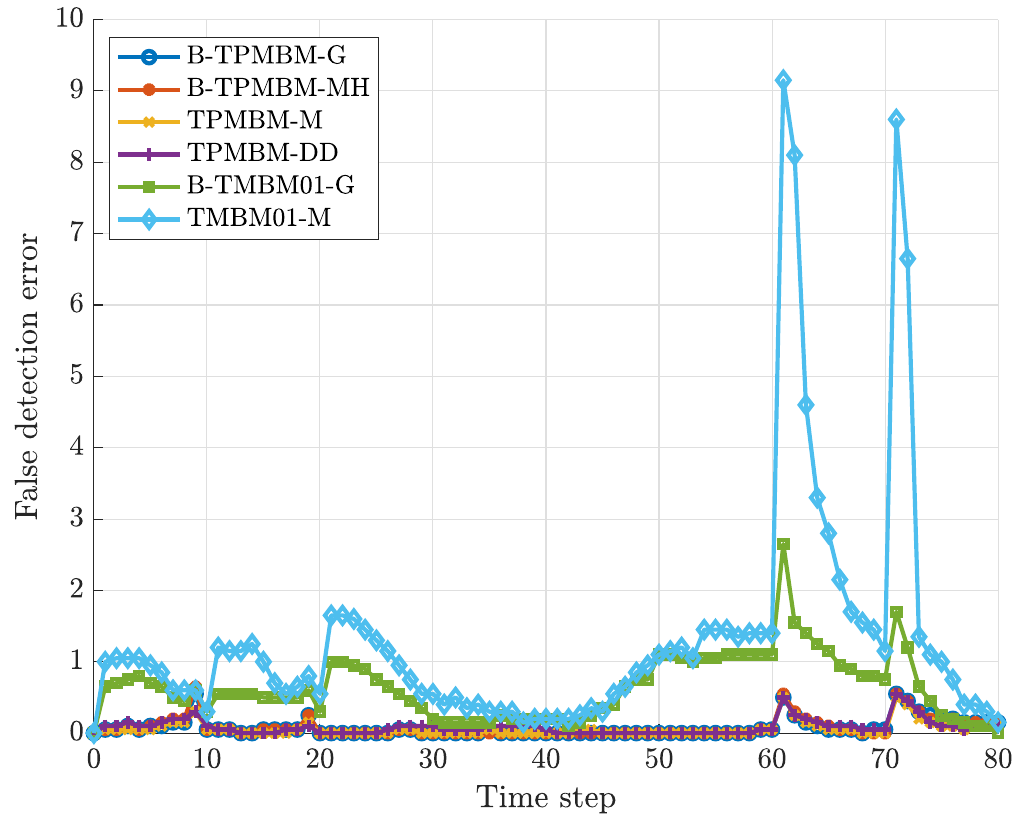}
  \includegraphics[width=0.24\linewidth]{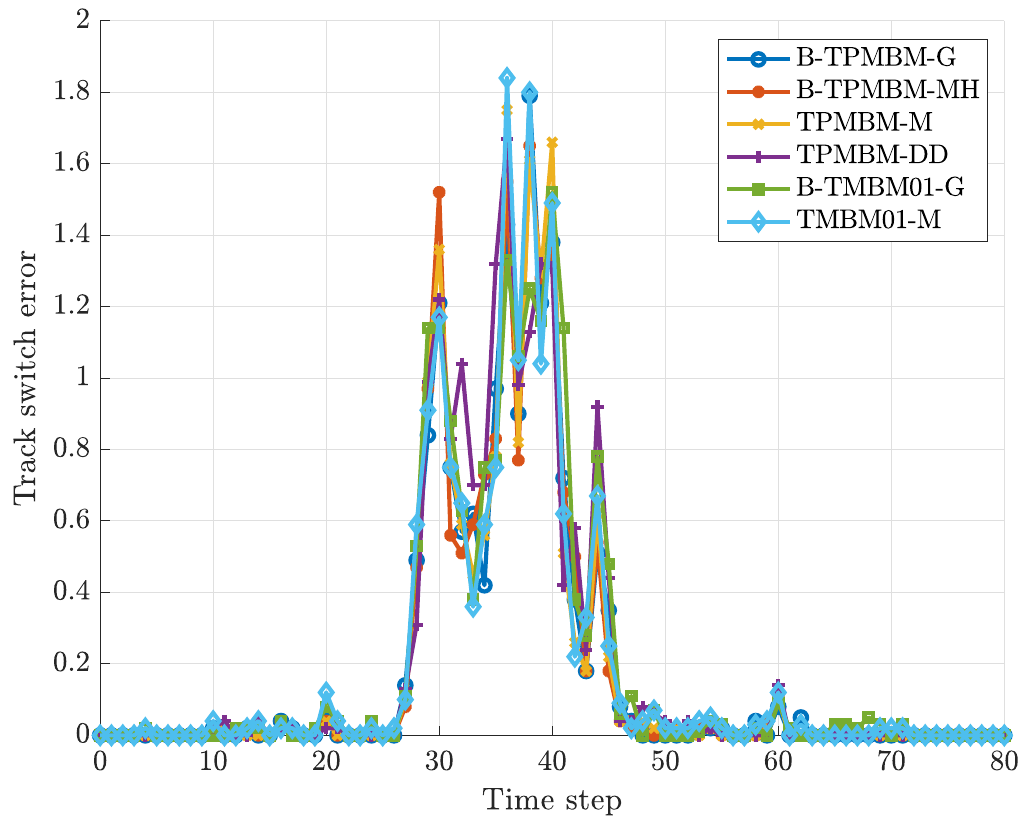}
  \caption{Trajectory GOSPA errors breakdown over time: The four figures arranged from left to right, respectively, illustrate the localization errors, missed detection errors, false detection errors, and track switch errors.}
  \label{fig_LPmetricDecom}
\end{figure*}

To further analyze the performance of different implementations, the trajectory GOSPA errors and their decompositions over time are shown in Fig. \ref{fig_LPmetric} and Fig. \ref{fig_LPmetricDecom}, respectively. The numerical values of these average trajectory GOSPA errors, together with the average runtime\footnote{MATLAB implementation on Apple M1 Pro.}, are presented in Table \ref{tab_LPmetric}. In Fig. \ref{fig_LPmetricDecom}, the localization error measures the total localization errors for all the detected objects, and it varies with the number of detected objects. The two figures showing the missed and false detection errors reflect the mismatch between the true and estimated number of objects. The peaks in these curves represent time steps when new objects are born or existing objects die, or objects move in proximity. In these cases, it is more difficult to detect and localize the objects. Finally, the peaks in the curves showing the track switch error correspond to the time steps when objects moving in proximity. In these cases, it is difficult to distinguish the case where objects are moving separately from the case where objects are moving cross each other, resulting in higher track switch errors.

From Fig. \ref{fig_LPmetric}, we can observe that on the whole B-TPMBM-MH presents the best trajectory estimation performance, especially fewer misdetections when objects are closely spaced. B-TPMBM-G has the second-best performance, outperforming TPMBM-M and TPMBM-DD by a margin. Also, $\text{B-TMBM}_{01}$-G outperforms T$\text{MBM}_{01}$-M by considering a multi-scan data association problem in a batch. Moreover, it is noteworthy that, despite being a batch method, $\text{B-TMBM}_{01}$-G presents worse estimation performance than TPMBM-M and T-PMBM-DD, mainly by showing larger false detection errors when objects die. This is because Bernoulli components in $\text{B-TMBM}_{01}$-G have deterministic existence, which results in an exponential increase of the number of global hypotheses in the prediction step. This makes it slower for MCMC sampling to explore the data association space. This result highlights the fact that TPMBM has a more efficient hypothesis structure, where object death events are captured in single-trajectory density via the pmf of trajectory end times.

Concerning the balance between computational complexity and estimation performance, TPMBM-M mainly depends on the maximum number of global hypotheses allowed in filtering recursion; TPMBM-DD mainly depends on the sliding window length; and B-TPMBM-G and B-TPMBM-MH mainly depend on the number of sampling iterations. The estimation performance of all the implementations could be further improved by increasing their computational budget. However, empirical results show that the performance gain from doing so is limited for online TPMBM implementations. In particular, we did not observe any performance improvement of TPMBM-DD by further increasing its sliding window length. As for TPMBM-M, although it is an online implementation, its runtime under the current setting is already comparable to B-TPMBM-MH. These analyses further highlight the importance of developing batch MOT implementations, which can break the bottleneck of the tracking performance of online implementations.

\begin{table}[!t]
  \caption{Performance of B-TPMBM-MH under different settings of the proposal distribution}
  \label{tab_ablation}
  \centering
  \begin{adjustbox}{width=1\linewidth}
  \begin{tabular}{ccc}
    \hline
    & Trajectory GOSPA & Runtime (s) \\ \hline
\begin{tabular}[c]{@{}c@{}}High track switch probability\\ $p(c=4)=1/2,p(c=1,2,3)=1/6$\end{tabular}   & 454.1            & 769         \\
\begin{tabular}[c]{@{}c@{}}Medium track switch probability\\ $p(c=1,4)=1/3,p(c=2,3)=1/6$\end{tabular} & 455.0            & 991         \\
\begin{tabular}[c]{@{}c@{}}Low track switch probability\\ $p(c=2,3,4)=1/6,p(c=1)=1/2$\end{tabular}    & 460.3            & 1201        \\ \hline
\end{tabular}
\end{adjustbox}
\end{table}

\begin{table}[!t]
  \caption{Trajectory GOSPA metric, its decomposition and runtime}
  \label{tab_LPmetric}
  \centering
  \begin{adjustbox}{width=1\linewidth}
  \begin{tabular}{ccccccc}
  \hline
  & Total & Localization & Missed & False & Switch & Runtime (s) \\ \hline
  TPMBM-M  & 477.7 & 341.2  & 116.5  & 4.6   & 15.5 & 613  \\
  TPMBM-DD  & 483.5 & 348.5  & 111.5  & 6.7   & 16.9  & 190  \\
  B-TPMBM-G  & 469.7 & 345.2  & 103.2   & 6.0  & 15.4  & 1241 \\
  B-TPMBM-MH  & \textbf{454.1} & 349.8  & 82.8   & 6.2   & 15.4 & 769  \\
  T$\text{MBM}_{01}$-M  & 597.0 & 358.3  & 115.4   & 107.1  & 16.2  &  410 \\
  $\text{B-TMBM}_{01}$-G & 541.5 & 349.5  & 123.1   & 52.3  & 16.2  & 1354 \\ \hline
  \end{tabular}
\end{adjustbox}
\end{table}

\section{Conclusions}

In this paper, we have presented two batch implementations of the TPMBM filter using Gibbs sampling and MH sampling, respectively. The simulation results demonstrate that the implementation using MH sampling has achieved state-of-the-art multi-trajectory estimation performance. This result is attributed to the efficient hypothesis structure of TPMBM, the trajectory density representation, and the careful design of the proposal distributions in MH sampling. 

An interesting follow-up work is to develop batch TPMBM implementations for extended object tracking. It would be also interesting to explore more advanced MCMC techniques \cite{luengo2020survey} for sampling the data associations. One promising sampling strategy is the use of multiple try Metropolis algorithm \cite{martino2018review}, where multiple samples are drawn from the proposal distribution at each iteration. Then, one of them is selected as a candidate for the next state according to some suitable weights. To adapt this idea to sample data associations, one may simultaneously consider track update move, merge/split move and track switch move at each iteration, fostering the exploration of the data association space.

MCMC sampling is a powerful technique to obtain samples from complex distributions, and in this work we use it to sample data associations. Another future line of research is to develop MCMC algorithms to obtain multiple samples from the posterior of the set of trajectories \cite{li2023adaptive}.

\bibliography{references.bib} 

\bibliographystyle{IEEEtran}

\begin{IEEEbiography}
  [{\includegraphics[width=1in,height=1.25in,clip,keepaspectratio]{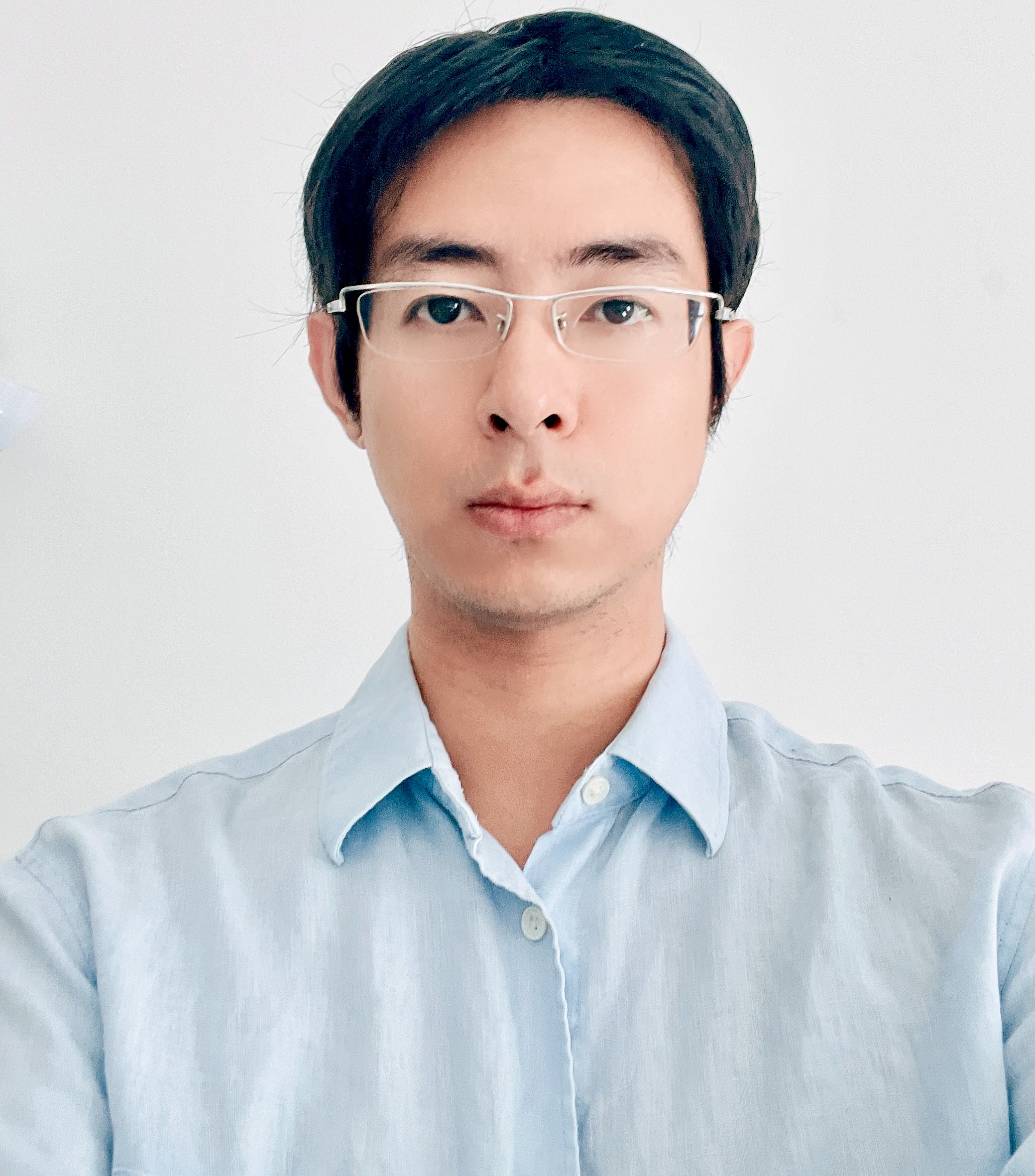}}]{Yuxuan Xia} received his M.Sc. degree in communication engineering and Ph.D. degree in signals and systems both from Chalmers University of Technology, Gothenburg, Sweden, in 2017 and 2022, respectively. After obtaining his Ph.D. degree, he stayed at the same research group as a postdoctoral researcher for a year. He is currently an industrial postdoctoral researcher with Zenseact AB, Gothenburg, Sweden and also affiliated with the Division of Automatic Control, Linköping University, Linköping, Sweden. His main research interests include sensor fusion, multi-object tracking and SLAM, especially for automotive applications. 
  \end{IEEEbiography}

\begin{IEEEbiography}
  [{\includegraphics[width=1in,height=1.25in,clip,keepaspectratio]{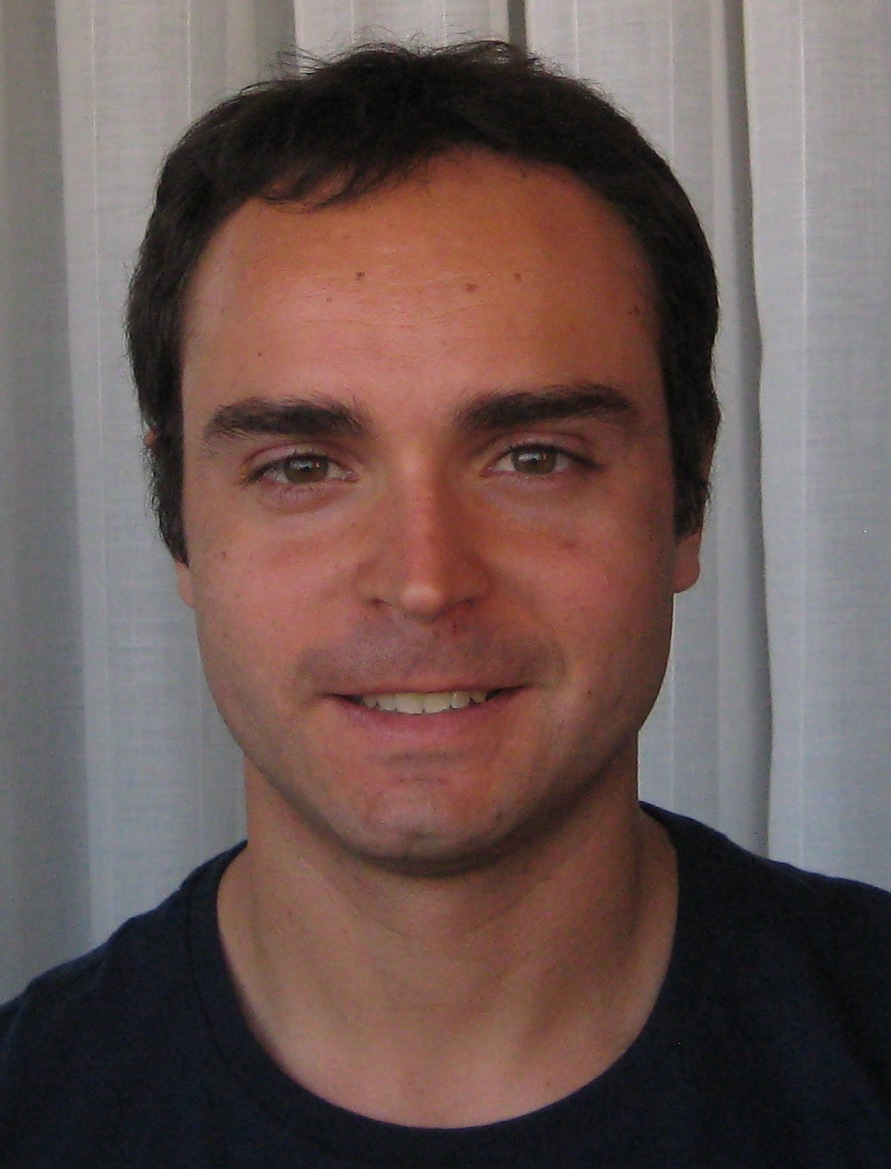}}]{{\'{A}}ngel F. Garc{\'{i}}a-Fern{\'{a}}ndez} received the telecommunication engineering degree and the Ph.D. degree from Universidad Politécnica de Madrid, Madrid, Spain, in 2007 and 2011, respectively. He is currently a Senior Lecturer in the Department of Electrical Engineering and Electronics at the University of Liverpool, Liverpool, UK. He previously held postdoctoral positions at Universidad Politécnica de Madrid, Chalmers University of Technology, Gothenburg, Sweden, Curtin University, Perth, Australia, and Aalto University, Espoo, Finland. His main research activities and interests are in the area of Bayesian inference, with emphasis on dynamic systems and multiple target tracking.
\end{IEEEbiography}

\begin{IEEEbiography}
  [{\includegraphics[width=1in,height=1.25in,clip,keepaspectratio]{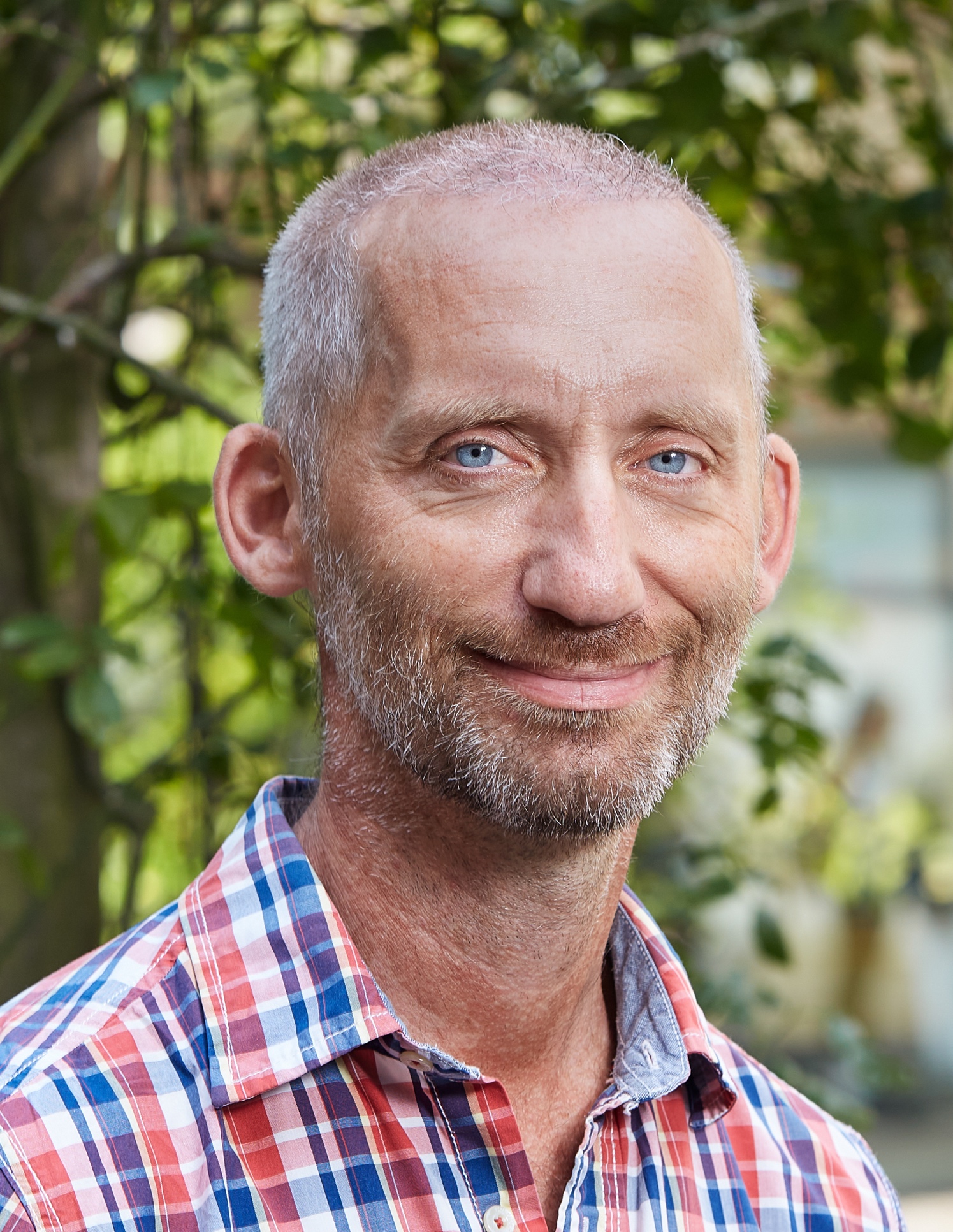}}]{Lennart Svensson} is a Professor of Signal Processing with the Chalmers University of Technology. His main research interests include machine learning and Bayesian inference in general, and nonlinear filtering, deep learning, and tracking in particular. He has organized a massive open online course on multiple object tracking, available on edX and YouTube, and received paper awards at the International Conference on Information Fusion in 2009, 2010, 2017, and 2019.
\end{IEEEbiography}

\end{document}